\documentclass[letterpaper,twocolumn,10pt]{article}
\usepackage{usenix2019_v3}

\usepackage{tikz}
\usepackage{amsmath}

\usepackage{tikz}
\usepackage{algorithm}
\usepackage{algpseudocode}
\usepackage{pifont}
\usepackage{booktabs}
\usepackage{listings}
\usepackage{enumitem}
\usepackage{multirow}
\usepackage{seqsplit}
\usepackage{msc}
\usepackage{color}
\usepackage{xurl}
\usepackage{comment}

\newcommand{\cmark}{\ding{51}} 
\newcommand{\xmark}{\ding{55}} 

\def\aname{E-Trojans}
\def\toolkit{\texttt{E-Trojans}}

\def\cuvt{1.58V}

\def\covt{4.7V}

\def\vulns{four}
\def\counters{four}

\def\vone{BCTRL firmware is unencrypted}
\def\cone{Encrypt the BCTRL firmware with TEA}

\def\vtwo{BCTRL firmware is unsigned}
\def\ctwo{Sign and verify the BCTRL firmware with ECDSA}

\def\vthree{UART bus lacks integrity protection, encryption, and authentication}
\def\cthree{Protect the UART bus with SCP03}

\def\vfour{UART bus lacks protection against DoS}
\def\cfour{Protect the UART bus with rate limiting}

\def\attacks{five}
\def\dos{eight} 

\def\aubr{Undervoltage Battery Ransomware}
\def\aubra{UBR}

\def\aobd{Overvoltage Battery Destruction}
\def\aobda{OBD}

\def\auti{User Tracking via Internals}
\def\autia{UTI}

\def\ades{Denial of E-Scooter Services}
\def\adesa{DES}

\def\aphl{Password Leak and Recovery}
\def\aphla{PLR}

\def\caps{eleven}

\begin{document}

\date{}

\title{\Large \bf \aname: Ransomware, Tracking, DoS, and Data Leaks on Battery-powered Embedded Systems}

\author{
{\rm Marco Casagrande}\\
KTH\\
casagr@kth.se
\and
{\rm Riccardo Cestaro}\\
University of Padua\\
riccardo.cestaro@outlook.it
\and
{\rm Mauro Conti}\\
University of Padua\\
mauro.conti@unipd.it
\and
{\rm Eleonora Losiouk}\\
University of Padua\\
eleonora.losiouk@unipd.it
\and
{\rm Daniele Antonioli}\\
EURECOM\\
daniele.antonioli@eurecom.fr
}

\maketitle

\begin{abstract}
Battery-powered embedded systems (BESs), such as smartphones, e-scooters, and drones, have become ubiquitous.
Their internals typically include a battery, a battery management system (BMS), a radio, and a motor controller.
Despite their safety implications, there is little research on the security and privacy of BES internals which are proprietary and undocumented.
To fill this gap, we present the first security and privacy assessment of the internals of two popular Xiaomi e-scooters.
We cover the M365 (2016) and Mi3 (2023) e-scooters and their Mi Home companion mobile app.
Via reverse-engineering (RE), we uncover \vulns\ critical design vulnerabilities, including a BMS remote code execution.

We develop \aname, \attacks\ novel attacks that flash a malicious BMS firmware over Bluetooth Low Energy (BLE)
from a malicious app on the victim's phone or within in BLE proximity.
\aname\ have a critical and real-world impact as they violate the safety, security, availability, and privacy of e-scooters and their users.
For instance, our battery ransomware permanently damages the battery using undervoltage and asks for a ransom payment,
while our user tracking attack uses unique e-scooter internal values a a fingerprint.
These attacks can be ported to other BESs with similar vulnerabilities.

We implement our attacks and RE findings in a modular and low-cost toolkit to test BES internals.
Our toolkit binary patches BMS firmware by adding malicious capabilities, such as disabling battery safety thresholds,
which enables our attacks and the creation of new ones.
We test our attacks on real M365 and Mi3 e-scooters, empirically confirming their effectiveness and practicality.
We propose \counters\ practical countermeasures to improve Xiaomi's security and privacy.
We responsibly disclosed our findings to Xiaomi, which acknowledged and addressed them.
\end{abstract}

\section{Introduction}\label{sec:intro}

Battery-powered embedded systems (BESs), such as electric cars, e-scooters, drones, and smartphones, are pervasive. Electric vehicles alone have a market size
of USD 623.5 billion~\cite{bev-market-size}. Meanwhile, \emph{e-scooters}, the focus of this work,
have a market of USD 37 billion and an estimated
annual growth of 9.9\%~\cite{escooter-market-size,escooter-market-size2}.
These devices carry sensitive data and can be remotely controlled with wireless signals.
Thus, vulnerable BES can be attacked to cause security, privacy, and safety issues.
For instance, a recent attack on pagers leveraged compromised internal components
to detonate explosives concealed within their batteries~\cite{pagers}.

E-scooters, and other BESs, have complex and proprietary \emph{internals} typically including a battery and its management system, an electric motor and its controller, and a wireless module supporting a low-power communication such as Bluetooth Low Energy (BLE). Moreover, the architecture includes internal buses, such as Universal Asynchronous Receiver-Transmitter (UART) and Inter-Integrated Circuit (I2C), allowing the modules to exchange data. An e-scooter is controlled by a user via a smartphone and a companion app.

BES internals are an attractive \emph{attack surface}, especially for remote attackers.
Automotive research discussed the security of externally reachable electric control units (ECUs)~\cite{checkoway2011comprehensive} and controller area network (CAN) buses~\cite{koscher2010experimental}.
Internal attack surfaces have been explored in works such as~\cite{wolf2004security},
implementing authentication and encrypted communication on automotive bus systems.
Drone literature highlighted issues in physical layer~\cite{drone-app-vuln} and communication protocols~\cite{drone-fwup,drone-bebop},
issues involving, among others, firmware updates and image transfer.

However, \emph{the security and privacy of e-scooters' internals} have received no attention from the research community. Researchers focused on external attack surfaces, including proprietary protocols over BLE~\cite{casagrande23espoofer} or studying the privacy of e-scooter rental mobile apps~\cite{scooter_wisec_2022} (refer to Section~\ref{subsec:mot} for an extended motivation).
We fill this gap by presenting a security and privacy assessment of the internals of two popular
e-scooters sold by Xiaomi and its subsidiary Segway-Ninebot, which are market leaders~\cite{xiaomi-leader}. We target the \emph{M365} and \emph{Mi3} models, covering two e-scooter generations from 2016 to 2024. These e-scooters are used by millions of people~\cite{ninebot-sold} who either own an e-scooter or rent it using services like Bird~\cite{bird-equals-m365,who-makes-bird}.
We also cover \emph{Mi Home}, the Xiaomi e-scooter companion app available for Android~\cite{mihome-android} and iOS~\cite{mihome-ios}.

We uncover \emph{\vulns\ design vulnerabilities} (V1--V4) affecting the M365 and Mi3 internals, including
remote code execution on the BMS via an unsigned firmware update, and unencrypted and unauthenticated internal communication. Being design flaws, they are effective on \emph{any} M365 and Mi3 e-scooter. Other e-scooters sharing the same internals, such as the Pro, Pro2, 1S, and Essential, are also vulnerable to the \aname\ attacks.
We discover the vulnerabilities by reverse-engineering the e-scooters' internals and reporting new findings, including low-level information about the battery management control and safety configuration.

We propose \textbf{\aname}, \emph{\attacks\ novel attacks on the e-scooter internals} taking advantage of V1--V4. The attacks force a firmware update over BLE to install malicious and unsigned BMS firmware to achieve different goals. For instance, we present an \emph{undervoltage battery ransomware} attack that irreversibly degrades the e-scooter's battery until the victim pays a ransom, even if the e-scooter is turned off and not plugged into a charger. We also present an \emph{overvoltage battery destruction} attack that permanently damages the battery while the e-scooter is connected to its charger. Both attacks can cause hazards, including fire and explosions due to physical phenomena like thermal runaway and polarity inversion~\cite{polarity-inversion}. Moreover, we show a novel way to track an e-scooter and its driver using a fingerprint computed from the e-scooter's internals (e.g., electric motor serial number) and advertising it over BLE.

\aname\ are conducted \emph{remotely} via a malicious app
installed on the victim's phone, or in \emph{wireless range}
of the e-scooter via a BLE device.
The attacks have a critical and large-scale impact on the Xiaomi e-scooter ecosystem as they violate the security, privacy, and safety of millions of e-scooters and their users. Moreover, they could be effective on other BESs (e.g., electric cars, drones, or smartphones), affected by issues like V1--V4.

We present \toolkit, a toolkit implementing the \aname\ attacks using low-cost and open-source software and hardware.
The toolkit contains a binary firmware patcher that generates malicious BMS firmware, allowing to deploy our attacks but also the creation of new ones by combining \caps\ capabilities, such as turning off firmware updates and safety-critical voltage protections.
Our toolkit targets STM8 chips, but can be extended to others, like STM32, with low engineering effort~\cite{stm8-migration-stm32}.

We confirm that the attacks are practical and effective by ethically testing them
on real and up-to-date 365 and Mi3 e-scooters in a controlled environment. Among others, we manage to overheat the Mi3 battery above 80°C  (i.e., the thermal stability threshold~\cite{overheat-stability}), degrade the M365 battery's autonomy by 50\% in less than four hours using an undervoltage battery ransomware,
and track the e-scooters using BLE.
To fix the presented vulnerabilities and attacks, we propose \emph{\counters\
countermeasures} (C1--C4). These add confidentiality and integrity
guarantees for the BMS (controller) firmware and protect the internal (UART)
bus against spoofing and DoS. Our fixes are \emph{backward-compatible} and \emph{low-cost},
as they reuse available cryptographic primitives and require minimal computational overhead.

We summarize our contributions as follows:
\begin{itemize}
  \item We identify the lack of security and privacy assessments on e-scooters' internals and present an analysis of two popular Xiaomi e-scooters (M365, Mi3). Via extensive RE, we uncover \vulns\ critical design vulnerabilities, including remote code execution on a BMS by flashing a malicious and unsigned firmware.
  \item We unveil \attacks\ novel attacks on e-scooter internals that we call \aname. The attacks are deployable remotely or in proximity and, among others, use battery overvoltage or undervoltage to damage the e-scooter battery and ask for a ransom or track an e-scooter via a fingerprint computed from the e-scooter's internal components.
  \item We release \toolkit, a toolkit implementing our attacks with open-source software and cheap hardware. The toolkit is usable to further RE the proprietary Xiaomi e-scooter ecosystem, target other BESs, and compose new attacks.
  \item We deploy our \attacks\ attacks on actual M365 and Mi3 e-scooters, empirically confirming their practicality and stealthiness. We propose \counters\ effective, low-cost, and legacy-compliant countermeasures to fix the \aname\ vulnerabilities and attacks.
\end{itemize}

\subsection*{Responsible Disclosure}

We followed standard practices for responsible disclosure by contacting Xiaomi and the manufacturers
of the BMS chips we target: STMicroelectronics (ST) and Texas Instruments (TI).
We notified Xiaomi in November 2023 via their official HackerOne bug bounty program~\cite{xiaomi-bugbounty}. Xiaomi assigned a Medium (5.8) CVSS severity score 
and closed our report as informative in March 2024.
In April 2024, we contacted TI and ST.
TI could not comment on the Xiaomi third-party firmware and highlighted that their STM8L151K6 chip
does not claim resistance to physical attacks (we note that we are not proposing physical attacks, but attacks that can be conducted in wireless proximity or remotely via a malicious mobile app).
ST did not respond to our report.

After a second round of disclosure in June 2025, Xiaomi acknowledged our attacks, awarding us with the highest bounty for our tier and a Medium severity CVE.
They also stated that the M365 and Mi3 models have reached the end of their lifecycle,
as described in~\cite{xiaomi-trust-center}, and that the \aname\ vulnerabilities have been mitigated
in all subsequent Xiaomi electric scooter models, which now incorporate enhanced security measures. Hence, our findings contributed to a more secure Xiaomi e-scooter ecosystem.

\section{Background, Motivation and Threat Model}\label{sec:mot-tm}

In this section, we provide background information, motivate our work, and present our system and attacker models.

\subsection{Background}\label{subsec:back-known}

\textbf{E-Scooter Block Diagram}.
As shown in Figure~\ref{fig:tm}, a Xiaomi e-scooter is composed of three systems:
\emph{Bluetooth Low Energy (BTS)}, \emph{Driving (DRV)}, and \emph{Battery Management (BMS)}.
BTS provides low-power and reliable wireless communication, DRV controls the electric motor, and BMS manages the battery.
Each system has a dedicated System-On-Chip (SoC) running closed-source and undocumented firmware.
The systems communicate using a proprietary protocol over a UART bus that provides a serial, asynchronous, and full-duplex digital communication channel with TX/RX wires.

\textbf{Mi Home App}.
The user interacts with a Xiaomi e-scooter via
the \emph{Mi Home} app for Android~\cite{mihome-android}
or iOS~\cite{mihome-ios}. The e-scooter and Mi Home discover each other using BLE scanning and advertising, and communicate using a proprietary protocol on top of BLE~\cite{casagrande23espoofer}.
The user logs in with their Xiaomi account on Mi Home,
pairs the app and the e-scooter, and then uses the app to interact with the e-scooter. Once paired, the devices automatically connect when they are in BLE range without requiring user interaction.
Mi Home offers valuable features for controlling the e-scooter, such as setting a password to lock it, performing over-the-air (OTA) firmware updates,
and monitoring the battery status.

\textbf{E-Scooter Internals}.
There is limited information about Xiaomi e-scooters' internals.
In~\cite{casagrande23espoofer}, the authors focus on external communications with Mi Home and do not discuss internal components.
Research on the Xiaomi 1S e-scooter reports that Xiaomi encrypts the BTS, DRV, and BMS binaries using TEA with a leaked encryption key~\cite{botox-tea}. These binaries are signed with ECDSA~\cite{robo-fw,botox-tea} and contain a certificate chain with a public key for signature verification.

\textbf{Firmware Update}.
During a firmware update, Mi Home transmits firmware to the BTS, which
decrypts, verifies, and distributes them to the DRV and the BMS over the UART bus.
The developers of the ScooterHacking Utility Android app
reversed an old Xiaomi firmware update protocol to enable flashing a modded
firmware on e-scooters~\cite{app-shu,home-shu}. However, there is no information about recent BTS, DRV, and BMS firmware and their update protocols.

\textbf{Battery}.
The M365 and Mi3 e-scooters are equipped with a lithium-ion (Li-Ion INR) battery pack.
These battery packs contain ten individual cells in a 5x2 configuration:
two parallel strings, each made of five cells connected in series.
Their nominal range goes from 42V (at 100\% charge) to 36V (at 0\% charge).

\subsection{\aname\ Motivation}\label{subsec:mot}

Vulnerabilities and attacks on BES internals are relevant, as they can cause security, privacy, and safety issues. For example, they can result in damage or destruction of physical components (e.g., the battery) or hazards (fire, explosion). 

Prior work in the area focused on automotive~\cite{kiley-tesla}, drones~\cite{drone-fwup, drone-bebop}, and laptops~\cite{miller2011battery}. No prior study analyzed \emph{internal attack surfaces for e-scooters}, and few papers looked at their external threats. In~\cite{casagrande23espoofer}, the authors reverse-engineered
and compromised proprietary communication protocols used by Xiaomi e-scooters and Mi Home over BLE.
E-scooters' security threats, such as DoS, physical attacks, and firmware vulnerabilities,
have been highlighted by~\cite{isik2023platforms}.
The authors of~\cite{cameron2019iot} performed a threat analysis on the Xiaomi M365
and reproduce known attacks on its authentication.
Privacy issues have also been found in e-scooters' apps~\cite{scooter_wisec_2022}
that shared user identity, location, and phone data with the manufacturer and third parties.
To address this gap, the paper explores the internal attack surface of an e-scooter and its (remote) exploitability.

\subsection{Threat Model}\label{subsec:tm}

\begin{figure}[tb]
    \centering
    \includegraphics[width=0.9\linewidth]{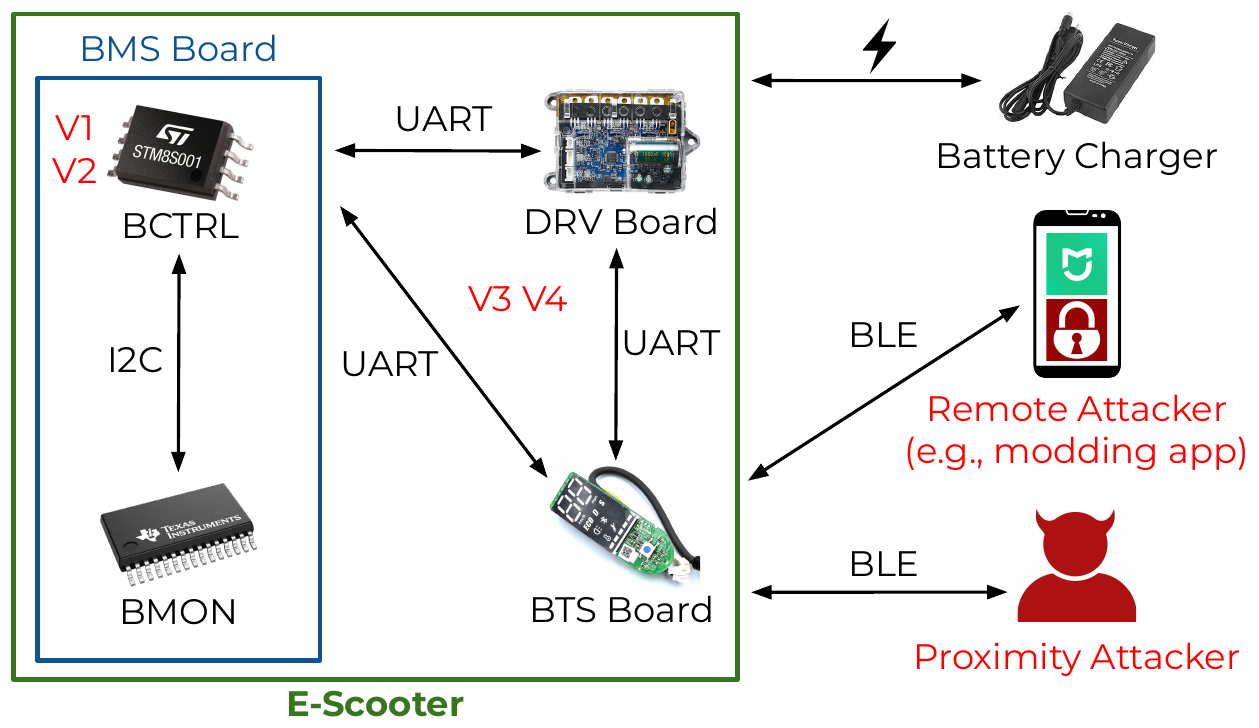}
    \caption{\aname\ block diagram and attacker models. The green rectangle shows e-scooter internals (BMS, DRV, and BTS boards, and UART and I2C buses). The blue rectangle shows BMS components (BCTRL and BMON connected via I2C). We consider a proximity attacker in the BLE range of the e-scooter and a remote one who installed a rogue app (e.g., an e-scooter modding app) on the victim's smartphone. We show four design vulnerabilities affecting the BCTRL (V1, V2) and the UART bus (V3, V4).}
    \label{fig:tm}
\end{figure}

\textbf{System Model}.
Our system model follows the block diagram in Figure~\ref{fig:tm} and includes an e-scooter, a charger, a smartphone, and a user. The e-scooter is either owned by the user or rented from a fleet and runs up-to-date firmware. The smartphone is owned by the user and runs Mi Home. The user paired their Mi Home app and account with the e-scooter. Moreover, they activate the software lock feature, allowing users to lock and unlock the e-scooter from the app.

\textbf{Attacker Model}.
We consider two attacker models: (i) a \emph{proximity-based attacker} who interacts with the e-scooter wirelessly over BLE (e.g., using their device in BLE range with the victim's e-scooter); and (ii) a \emph{remote attacker} who connects to the victim's e-scooter over the Internet through a malicious app installed on the victim's smartphone. The attackers are shown in red on the bottom right of Figure~\ref{fig:tm}. The proximity-based and remote attacker models map to real-world attack scenarios. For example, the proximity-based attacker can place a BLE device in a train station or a parking spot to target e-scooters in range. The remote attacker can install a malicious app on a victim's smartphone.

The attacker's main goal is to violate the \emph{safety}, \emph{security}, \emph{privacy},
and \emph{availability} of the e-scooter and its driver by attacking its internal components. For example, the attacker might want to attack the BMS or the DRV subsystems via a malicious firmware update over BLE.
Achieving these goals leads to a large-scale and critical impact on the Xiaomi e-scooter ecosystem by affecting millions of e-scooters and their users.
The attacker reuses tools, knowledge, and techniques from prior work~\cite{casagrande23espoofer,app-shu,robo-fw}. For example, they can bypass e-scooter authentication mechanisms over BLE using \texttt{E-Spoofer}~\cite{esp-gh}. The attacker \emph{has no physical access} to the victim's e-scooter, charger, and smartphone.

\section{Xiaomi E-Scooters RE and Vulnerabilities}\label{sec:back}

This section summarizes our RE findings on the M365 and Mi3 with an emphasis on their BMS. Then, it describes the \vulns\ \aname\ vulnerabilities.

\subsection{RE Findings}\label{subsec:back-unknown}

\begin{figure}[tb]
    \centering
    \includegraphics[width=0.7\linewidth]{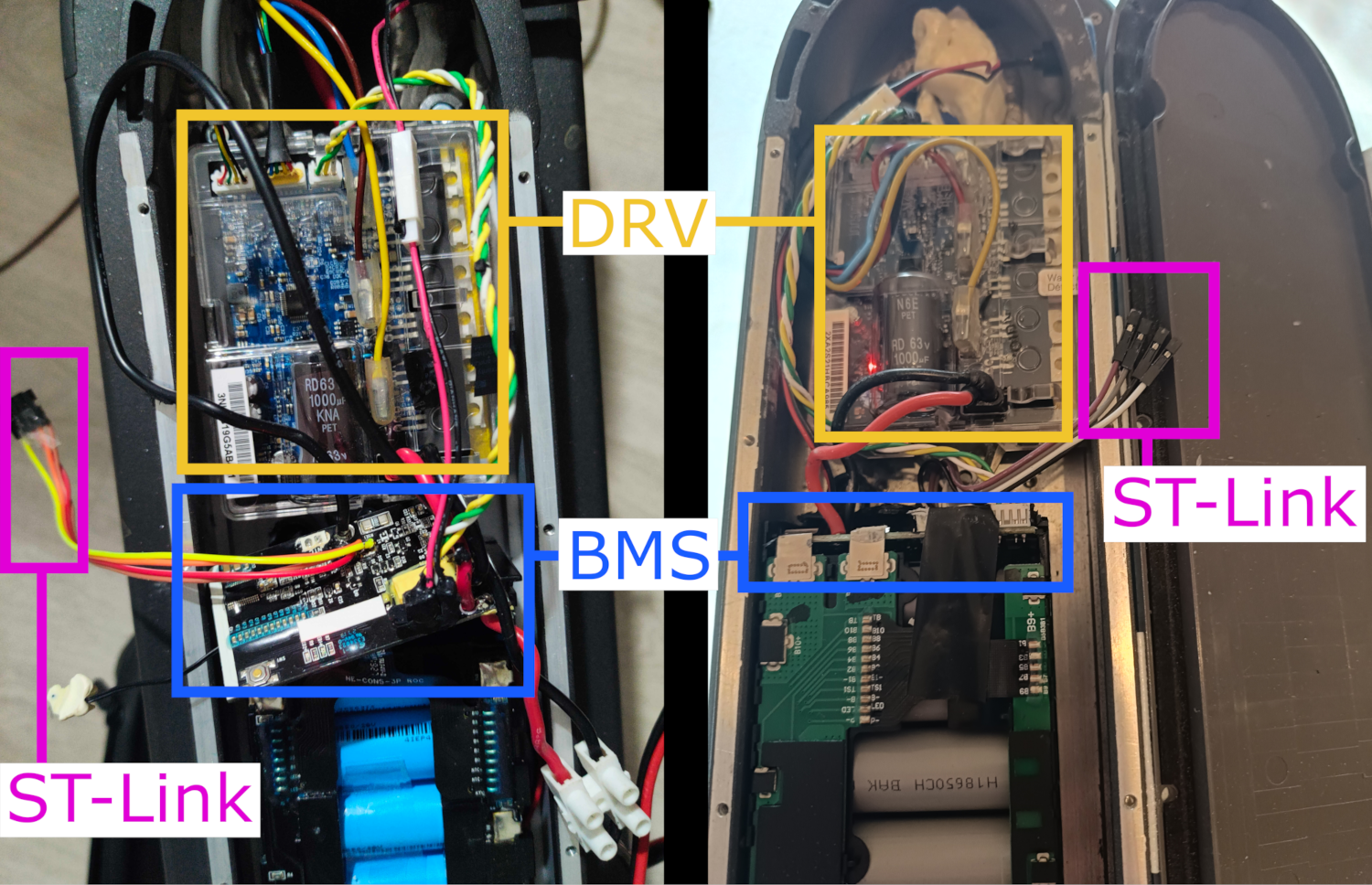}
    \caption{Disassembled M365 (left) and Mi3 (right). We color-coded the boxes
    to visualize the DRV (orange), the BMS (blue), and the soldered ST-Link wires (fuchsia).}
    \label{fig:disass}
\end{figure}

In 2023, we ran an extensive RE effort on the Mi3 and M365 e-scooters. At that time, we picked these two e-scooters because the Mi3 was the latest e-scooter from Xiaomi, and the M365 was the most popular one. These models represent two different e-scooter generations. However, they share the same internal architecture, BCTRL firmware, and communication protocols over UART and I2C.

We disassembled the M365 and Mi3 e-scooters by tearing down their lower deck,
gaining access to the DRV, BMS, UART bus, and battery pack, as shown in Figure~\ref{fig:disass}.
We extracted and inspected the BMS board of the e-scooters, finding a \emph{battery controller (BCTRL)} and a \emph{battery monitor (BMON)}.
The BCTRL and BMON are interconnected via an I2C bus, which is a synchronous two-wire serial interface (SDA and SCL) used for communication between integrated circuits.

Next, we detail new RE findings on the BCTRL and BMON of the Mi3 and M365, as they are needed to describe the \aname\ issues and the attacks.  These components work together to ensure safe battery management.
Our analysis focuses on three unsafe conditions that impact battery health and user safety: \emph{overvoltage (OV)}, which can lead to overheating, fire, and thermal runaway; \emph{undervoltage (UV)}, which can lead to polarity inversion and permanent battery damage; and voltage \emph{imbalance} between battery cells, which contributes to both undervoltage and overvoltage, accelerating battery degradation~\cite{mechanism-uv,battery-ov}.
Other RE findings, including details about the BTS and DRV firmware, and the proprietary UART and I2C protocols, can be found in~\cite{casagrande23espoofer,website-scooterhacking,botox-fwpatcher}.

\textbf{BMON}.
The BMON is a TI BQ76930 chip~\cite{BQ76930}. It measures battery parameters (i.e., voltage and temperature) via dedicated sensors, reports faults to the BCTRL, and stores safety thresholds (i.e., OV and UV thresholds). These thresholds are kept in dedicated BMON registers and are configurable by the BCTRL with read and write commands sent using proprietary I2C messages. When a safety threshold is surpassed, the BMON relies on hardware protections, such as a hardware load balancer for the battery cells, or charge and discharge MOSFETs to stop charging and discharging. These protections are used to prevent or mitigate hazardous situations, like battery overvoltage or undervoltage. A production BMON does not support firmware updates or hardware debugging~\cite{BQ-datasheet}.

\textbf{BCTRL}.
The BCTRL is an ST STM8L151K6 chip~\cite{STM8}. It manages the battery’s health with the BMON and communicates with the DRV and BTS. For example, it powers off the DRV and BTS when the battery undervolts or overvolts, initializes BMON registers during startup, and processes battery data received from the BMON. 
We identified the power supply (VDD), Single Wire Interface Module (SWIM), Ground (GND), and Reset (RST) pins used for powering, programming, and resetting the BCTRL. We soldered them to an \emph{ST-Link v2} programmer for firmware extraction and debugging, finding \emph{no hardware debug (ST-Link) protection}.

We RE the BCTRL firmware initialization and main loop by decompiling its stripped binary and debugging it via ST-Link. During initialization, the BCTRL loads its default settings, including serial number and firmware version. It also configures the BMON over I2C, including its overvoltage and undervoltage thresholds and the voltage balancing delta.
In the main loop, the BCTRL firmware: communicates over UART with the BTS and DRV, reads the voltage of individual battery cells from BMON read-only registers, and checks these voltages against the overvoltage and undervoltage thresholds and the voltage balancing delta. If a safety check fails, the BCTRL sends an OV or UV fault message over UART, which powers off the BTS and DRV, enters a low energy consumption mode, and performs voltage load balancing on imbalanced cells.

\subsection{\aname\ Vulnerabilities}\label{subsec:vulns}

Through RE, we uncovered \emph{\vulns\ design vulnerabilities} on the M365 and Mi3 internals:
\begin{enumerate}
  \item[\textbf{V1:}] \textbf{\vone}. An attacker can retrieve the unencrypted BCTRL firmware from Mi Home (at rest) or during a BLE firmware update (in transit).

  \item[\textbf{V2:}] \textbf{\vtwo}. An attacker can modify a BCTRL firmware or craft a new one,
  and flash it on the BCTRL without having to authenticate it.

  \item[\textbf{V3:}] \textbf{\vthree}. An attacker on the UART bus can eavesdrop on the UART messages,
  replay or forge them, and impersonate or MitM the DRV, BMS, and BTS.

  \item[\textbf{V4:}] \textbf{\vfour}. An attacker accessing the UART bus
  can DoS the BMS, BTS, and DRV.
\end{enumerate}

These issues stem from \emph{design flaws} in Xiaomi's firmware protection mechanisms and proprietary protocols and are independent from the hardware and software implementation details of the e-scooters. As a result, our vulnerabilities impact all M365 and Mi3 e-scooters and other e-scooters using the same vulnerable internals.

\section{\aname\ Attacks}\label{sec:attacks}
We present \emph{\aname, \attacks\ novel attacks on the internals of Xiaomi e-scooters}: \aobd\ (\aobda), \aubr\ (\aubra), \auti\ (\autia), \ades\ (\adesa), and \aphl\ (\aphla). These attacks can progressively and irreversibly damage the e-scooter battery via overvoltage and undervoltage, while asking for a ransom, track the e-scooter via BLE using internal fingerprints, and exfiltrate confidential data. They take advantage of V1--V4 discussed in Section~\ref{subsec:vulns}. These vulnerabilities exist in the M365, Mi3, and all other models that share the same BCTRL firmware and internal architecture, such as the Pro, Pro2, 1S, and Essential. As such, our attacks affect a wide range of e-scooters.

\begin{figure}[tb]
    \centering
    \includegraphics[width=\linewidth]{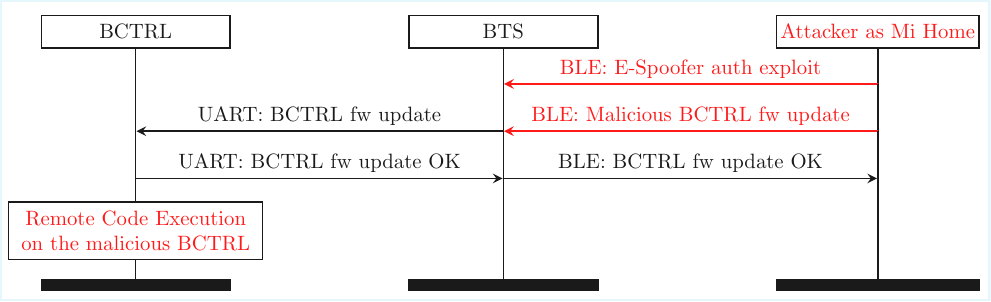}
    \caption{\aname\ attack technique. The attacker, impersonating Mi Home, utilizes the Malicious Pairing or Session Downgrade technique from E-Spoofer~\cite{esp-gh} to authenticate to the BTS of the victim's e-scooter. Then, they perform a rogue firmware update that installs a malicious BCTRL, resulting in remote code execution.}
    \label{fig:atk-tech}
\end{figure}

The \aname\ attacks flash a malicious and unsigned BCTRL firmware onto the victim's e-scooter via BLE.
The rogue firmware update can be accomplished in multiple ways. For example, Figure~\ref{fig:atk-tech} shows how to break authentication in the Xiaomi BLE proprietary protocol using tools such as the \texttt{E-Spoofer} toolkit~\cite{esp-gh}. Other remote attack vectors can tamper with the unsigned BCTRL firmware, like a software supply chain attack on the Mi Home firmware distribution server.

Our attacks raise severe safety concerns, including the risk of fires and explosions, by overvolting and undervolting the e-scooter battery beyond the maximum overvoltage threshold or the minimum undervoltage threshold. We evaluate such safety hazards in Section~\ref{subsec:safety}. Next, we describe each attack and its impact.

\subsection{\aobd\ (\aobda)}\label{subsec:aobda}

\aobda\ exploits overvoltage to damage and destroy the e-scooter battery, resulting in a DoS attack, which also threatens the user's safety.

\textbf{Battery Overvoltage}.
A fully charged cell in a Xiaomi e-scooter battery pack has 4.2V nominal voltage.
This is also the default value of the \emph{overvoltage threshold},
stored in the BMON's \texttt{OV\_TRIP} register and used to protect the battery.
Overvoltage occurs when the voltage of a cell exceeds the overvoltage threshold and goes
above 100\% charging state. This is because 0\% and 100\% are relative values depending on
the chemistry of the battery and not absolute ones.
An overvolted cell is a serious safety hazard, capable of causing
permanent battery damage, cell voltage imbalance, battery overheating, and even fire or an explosion~\cite{battery-ov}. In case of overvoltage, the BMON raises a fault (OV fault bit =1, \texttt{sys\_stat} register) that can be cleared (bit set back to 0) by the BCTRL.

\begin{figure}[tb]
    \centering
    \includegraphics[width=\linewidth]{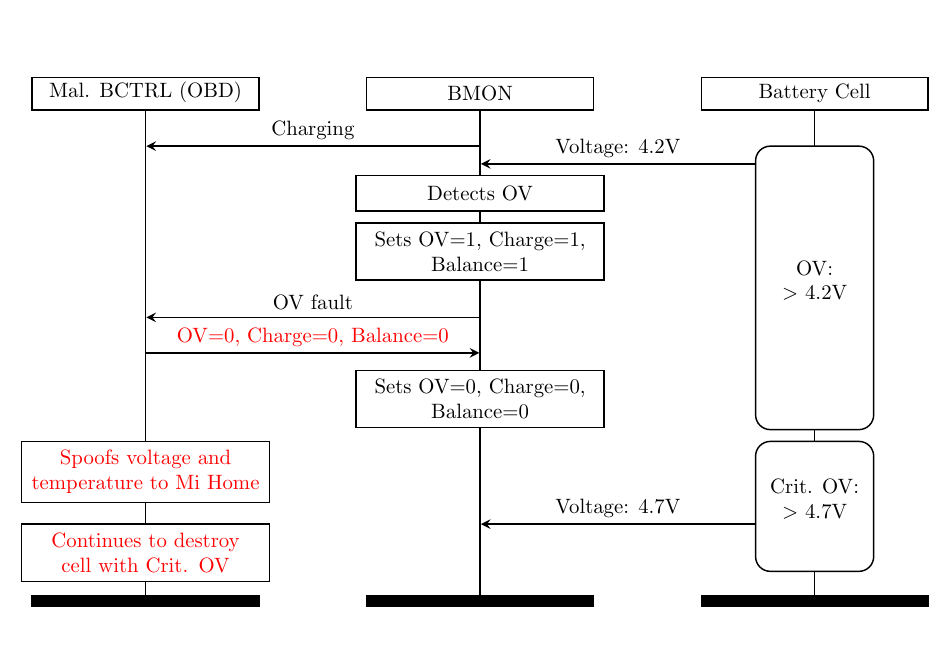}
    \caption{\aobd\ (\aobda). The Malicious BCTRL (the attacker) sets the OV threshold
    to the highest value (i.e., 4.7V). When the BMON detects overvoltage
    and tries to mitigate it, the attacker stops it by modifying the overvoltage fault bit,
    the MOSFET charge bit, and the voltage balancing bits.
    It also maintains stealthiness by spoofing fake sensor readings to Mi Home.}
    \label{fig:obd}
\end{figure}

\textbf{Attack Description}.
\aobda\ employs a malicious BCTRL firmware to disable overvoltage and voltage balancing protections in the BMON, overvolting and ultimately destroying the e-scooter battery cells, as shown in Figure~\ref{fig:obd}.
\aobda\ activates when the battery is charging. Upon reaching the default undervoltage threshold of 4.2V, the BMON overvoltage safety protections are triggered (e.g., OV fault and stop charging). However, \aobda\ defeats them by clearing the OV fault (OV fault bit =0, \texttt{sys\_stat} register), keeping the charger connected to the battery (MOSFET charge bit =0, \texttt{sys\_stat} register), and disabling load balancing (voltage balancing bits =0, \texttt{CELLBAL} registers). This technique enables what we define as \emph{critical overvoltage} (i.e., voltage above 4.7V). \aobda\ remains undetected by spoofing voltage and temperature measurements to Mi Home. Since our firmware disables voltage balancing, the battery cells remain in an overvolted and unbalanced state that degrades and destroys them over time. Unplugging the charger does not stop the damage, but prevents further overvoltage.

\textbf{Impact}. \aobda\ reveals, for the first time, a \emph{software-induced safety attack} on BESs, performed by a malicious BCTRL that can cause critical overvoltage without physical access or custom hardware.

\subsection{\aubr\ (\aubra)}\label{subsec:aubra}

\aubra\ is an \emph{undervoltage ransomware} that progressively and irreversibly damages the e-scooter's battery.
As opposed to classical ransomware, which targets files, it asks for a ransom to stop ongoing physical damage to the e-scooter battery.
It can also be deployed as a DoS attack that undervolts and destroys the battery without asking for a ransom.

\textbf{Battery Undervoltage}.
An e-scooter battery cell is considered discharged (0\% charge) when its voltage is 3.6V, the nominal value defined by the manufacturer.
However, cell voltage can drop as low as 2.75V, which is the default \emph{undervoltage threshold}, below which the cell is considered undervolted.
This threshold is stored in the BMON's \texttt{UV\_TRIP} register and applies to all battery cells.
Undervoltage reduces battery longevity and health~\cite{battery-uv,ti-ocuv,mechanism-uv},
prevents safe recharging, and can result in safety hazards. For example, it can trigger polarity inversion, which can cause safety risks such as internal short circuits, voltage imbalance, swelling, fires, or an explosion. In case of undervoltage, the BMON raises a fault (UV fault bit =1, \texttt{sys\_stat} register) and the fault can be cleared (bit set back to 0) by the BCTRL.

\begin{figure}[tb]
    \centering
    \includegraphics[width=\linewidth]{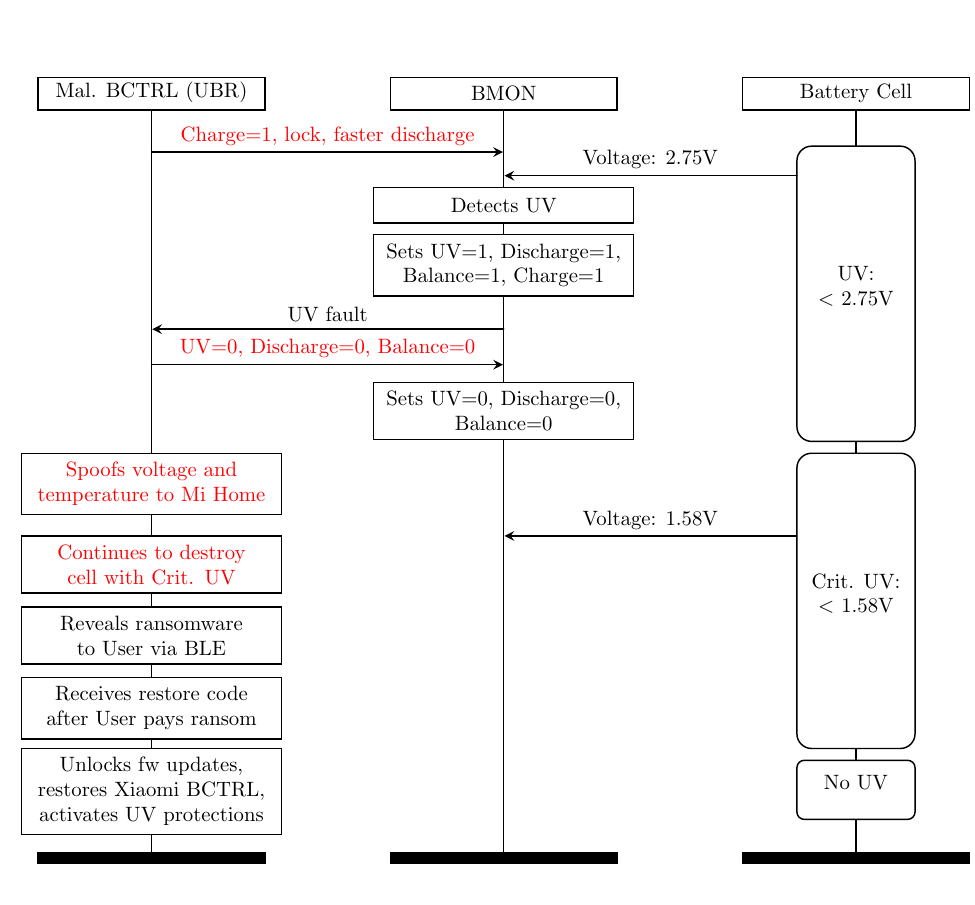}
    \caption{\aubr\ (\aubra). The Malicious BCTRL (the attacker) sets the UV threshold
    to the lowest value (i.e., 1.58V), locks the escooter, and prevents it from charging (MOSFET charge bit).
    When the BMON detects undervoltage, the attacker breaks safety protections
    by altering the undervoltage fault bit,
    the MOSFET discharge bit, and the voltage balancing bits.
    \aubra\ maintains stealthiness by spoofing sensor data, but, when triggered,
    will reveal its presence via BLE advertising.
    If the user pays the ransom, the ransom payment app submits the restore code
    that enables firmware updates and flashes the legitimate BCTRL firmware.}
    \label{fig:ubr}
\end{figure}

\textbf{Attack Description}.
\aubra\ disables the BMON undervoltage and voltage balancing protections to undervolt the e-scooter battery cells, as shown in Figure~\ref{fig:ubr}.
First, \aubra\ prevents charging (MOSFET charge bit =1, \texttt{sys\_stat} register), locks the e-scooter to keep it powered on, and forces discharging by ignoring sleep instructions and, optionally, turning on the headlight and taillight. When cell voltage reaches 2.75V, the BMON undervoltage safety protections activate (e.g., UV fault and voltage balancing). However, \aubra\ turns them off by clearing the UV fault (UV fault bit =0, \texttt{sys\_stat} register), restoring current flow from the battery to other internals (MOSFET discharge bit =0, \texttt{sys\_stat} register), and disabling voltage load balancing. This technique allows voltage to drop below 1.58V, putting cells into a condition that we call \emph{critical undervoltage}. The ransomware maintains stealthiness by spoofing safe battery parameters to Mi Home until a programmable trigger activates.
\aubra\ then reveals the ransomware to the user, rebooting the e-scooter to broadcast a short link via BLE advertising (e.g., \emph{t.ly/AaBbCc}). This link enables the victim to download an app to pay the ransom and see real-time battery degradation. To stop the damage, the victim has to transfer cryptocurrency to the attacker's crypto wallet. Once the attacker receives the payment, they share with the user through the ransom app a restore code unique to that e-scooter. The ransom app sends a special BLE packet containing that code to the e-scooter and then restores a legitimate BCTRL firmware. Undervoltage is fixed by reactivating the safety protections and charging the battery to a safe level.

\textbf{Impact}. \aubra\ introduces the \emph{first persistent, hardware-level ransomware} that monetizes physical degradation instead of data encryption. Attackers can rely on firmware vulnerabilities to cause irreversible hardware damage for profit.

\subsection{\auti\ (\autia)}\label{subsec:autia}

\autia\ is a privacy attack that tracks an e-scooter and its driver via a hardware-based fingerprint computed from the e-scooter's internals. Moreover, it can be used to leak the driver's data, such as travel mileage.

\begin{figure}[tb]
    \centering
    \includegraphics[width=\linewidth]{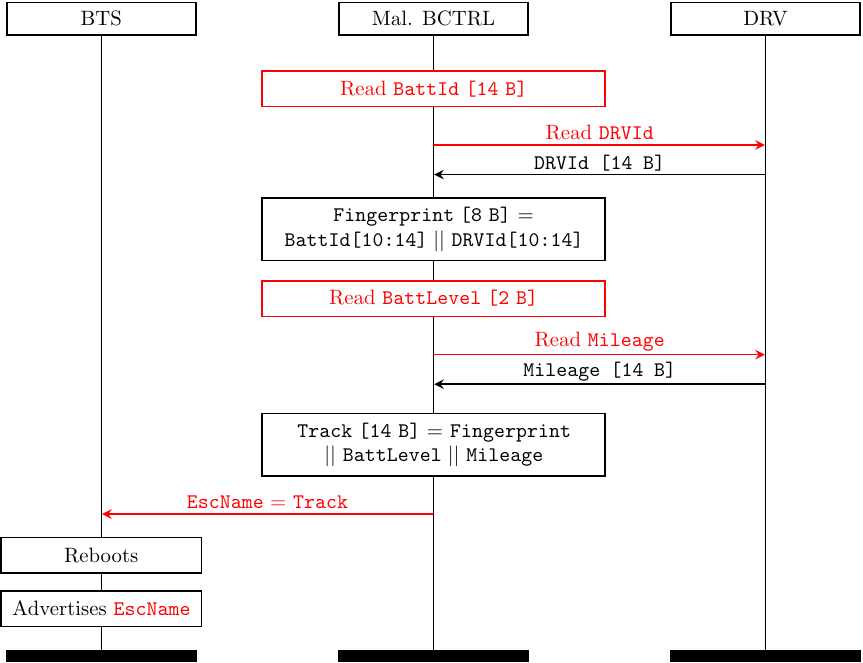}
    \caption{\auti\ (\autia).
        The Malicious BCTRL (the attacker) builds an 8-byte e-scooter fingerprint (\texttt{Fingerprint})
        from the electric motor serial number. Then, they append other sensitive data
        like mileage and battery level to the fingerprint, creating a 14-byte tracking message (\texttt{Track}).
        Finally, they enforce a BLE name change to \texttt{Track}, allowing anyone to track
        the user and read their private data in real-time.}
    \label{msc:attack-track}
\end{figure}

\textbf{Attack Description}.
Figure~\ref{msc:attack-track} shows how \autia\ tracks users and leaks their e-scooter data.
First, \autia\ reads the 14-byte battery serial number (\texttt{BattId}),
and retrieves the 14-byte DRV serial number (Read \texttt{DRVId})
by impersonating the BTS and spoofing a read operation to the DRV.
The malicious BCTRL, controlled by the attacker,
builds a \texttt{Fingerprint} from the last 8 bytes of \texttt{DRVId} representing the year,
month, and day of production, the revision, and the unit identifier (i.e., the nth unit produced that day).
\autia\ optionally retrieves via UART or I2C extra private data from the DRV, BMON, and BTS,
like the real-time mileage and battery level (Read \texttt{BattLevel} and Read \texttt{Mileage}).
Then, \autia\ appends the data to the fingerprint and creates a 14-byte tracking message (\texttt{Track}).
It sends a command to the BTS while spoofing the DRV to set the tracking message as the BLE name (\texttt{EscName = Track}).
The e-scooter reboots and advertises its new name (\texttt{Track}) over BLE.

\textbf{Impact}. \autia\ unveils a new \emph{hardware-fingerprinting vector} based on internal serial numbers and telemetry leaked through manipulated BLE advertisements. Unlike traditional BLE-address-based tracking, this approach persists across factory resets and BLE randomization. The attack demonstrates that DRV internal identifiers can uniquely and permanently tag devices, introducing a previously unknown privacy risk.

\subsection{\ades\ (\adesa)}\label{subsec:adesa}

\adesa\ is a collection of \dos\ DoS attacks (DES1--DES8). For instance, DES7 is a \emph{denial-of-sleep (DoSL)} attack which prevents the e-scooter from going to sleep, in order to deplete its battery~\cite{uher2016denial}. DES8 tampers with regenerative braking to injure the user by suddenly and unpredictably decelerating while they are driving down a slope. 

The \adesa\ attacks are detailed as follows: 
\begin{enumerate}
\item[DES1:] drops all UART and I2C packets sent to the BCTRL.
\item[DES2:] forces a supplier-only mode (SHIP mode) that disables the BMON by issuing a proprietary I2C command.
\item[DES3:] floods the UART bus with dummy packets.
\item[DES4:] periodically issues commands that reboot the e-scooter, causing a bootloop, or lock its motor, preventing the user from unlocking it.
\item[DES5:] disables battery charging unsetting the \texttt{canCharge} flag in the BCTRL flash.
\item[DES6:] exploits error handling and raises fake errors that cause the e-scooter to lock the motor or become unresponsive.
\item[DES7:] is a DoSL attack that keeps the BCTRL in a resource-intensive mode (other than low-energy mode) to discharge the battery faster.
\item[DES8:] alters braking settings and turns on/off regenerative braking (i.e., BCTRL slowing the e-scooter by converting kinetic energy into battery power, typically downhill).
\end{enumerate}

\textbf{Impact}. \adesa\ expands the concept of DoS beyond network availability to embedded denial. By exploiting internal buses and control logic, it shows that low-cost firmware faults can disable propulsion, braking, charging, or entire subsystems. This attack highlights the need to examine \emph{internal system communication (UART/I2C) as a DoS surface}.

\subsection{\aphl\ (\aphla)}\label{subsec:aphla}

\aphla\ enables recovering the Mi Home password set by the victim user by leaking its hash from the e-scooter memory and brute forcing it offline.

\textbf{Attack Description}.
On the Mi Home app, the user can set a password to prevent unauthorized access to their e-scooter via their smartphone. The password is stored on the phone, but we realized that its SHA256 hash is also stored in the DRV memory and is readable by the BTS via UART.
\aphla\ (BCTRL) spoofs the BTS to retrieve the password hash from the DRV,
sets the e-scooter BLE name equal to the hash, and exfiltrates it using the e-scooter BLE name via BLE advertising.
\aphla\ cycles through three BLE 14-byte BLE names containing portions of the hash. Each time it changes the BLE name, the e-scooter reboots. The password is six digits long (e.g., \texttt{123456}), and its search space is $10^6$~\cite{markert2020pin}. The attacker reconstructs the hash from the BLE advertisements and uses techniques such as password lists and rainbow tables to brute force the password offline.

\textbf{Impact}. \aphla\ exposes a new leakage channel involving a secret stored in the DRV that can be exfiltrated over BLE advertising. This highlights how \emph{weak separation between subsystems allows recovery of authentication information}, underlining the need for secure firmware boundaries and encrypted internal buses.

\subsection{Mapping Attacks and Vulnerabilities}

\begin{table}[tb]
  \caption{Mapping between the \attacks\ attacks and the \vulns\ vulnerabilities.
  All attacks exploit V1, V2, and V3. Only the DES attacks exploit V4.}
  \renewcommand{\arraystretch}{1.1}
  \centering\small
    \begin{tabular}{@{}lcccc@{}}
	    \toprule
	    \textbf{\aname\ Attack} & \textbf{V1} & \textbf{V2} & \textbf{V3} & \textbf{V4} \\
	    \midrule
           Overvoltage Battery Overheating (\aobda) &  \cmark\ & \cmark\ & \cmark\ & \xmark\ \\
          Undervoltage Battery Ransomware (\aubra) &  \cmark\ & \cmark\ & \cmark\ & \xmark\ \\ 
          User Tracking via Internals (\autia)  & \cmark\ & \cmark\ & \cmark\  & \xmark\ \\
	    Denial of E-Scooter Services (\adesa)  & \cmark\ & \cmark\ & \cmark\ & \cmark\ \\
	    Password Leak and Recovery (\aphla)  & \cmark\ & \cmark\ & \cmark\ & \xmark\ \\
	    \bottomrule
    \end{tabular}
    \label{tab:vulns}
\end{table}

Table~\ref{tab:vulns} shows the mapping between the \aname\ attacks and vulnerabilities.
The attacks take advantage of unencrypted BCTRL firmware (V1) as they rely on functionalities
that we RE from unencrypted firmware. They exploit the unsigned BCTRL firmware (V2) to push a malicious firmware on the BCTRL without integrity protection.
They also abuse the unprotected UART bus (V3) as they utilize malicious UART messages to impersonate internal components. Finally, \adesa\ exploits the lack of DoS protection on the UART bus (V4)
to DoS its internals.

\section{Implementation}\label{sec:impl}

We present \toolkit, a new toolkit to conduct the \aname\ attacks and create new ones. It includes code to send a rogue BCTRL firmware update and a binary patcher to build a firmware based on eleven malicious capabilities. While our attacks utilize fixed combinations (e.g., UTI uses C1, C2, C3, and C9), these eleven capabilities can be recombined to generate new attacks. The toolkit can be extended to work with other devices with internals and chips similar to the M365 and Mi3 e-scooters.

\subsection{BCTRL Binary Patching and Malicious Capabilities}\label{subsec:impl-capabilities}

The toolkit includes \texttt{payload-patcher.py}, a Python 3 script to binary patch a BCTRL firmware using STM8 assembly code as its input. Our script patches a vanilla BCTRL firmware, updating its CRC (used for integrity protection), and outputs a binary BCTRL file that can be flashed over BLE.
Moreover, \texttt{payload-patcher.py} offers an interactive shell to craft malicious BCTRL firmware. The user must select one of our \attacks\ attacks and a supported e-scooter model (i.e., M365, Pro, Essential, Pro2, and Mi3). We support models \emph{outside} of the M365 and Mi3 because they share the same BCTRL firmware and internal architecture, so our malicious BCTRL and rogue firmware update remain effective.

The binary patcher includes \emph{\caps\ capabilities (C1--C11)} that can be toggled and combined to test the \aname\ attacks or create new attacks.
\begin{itemize}
    \item[C1:] \emph{Disable Firmware Updates (DFU)}. DFU disables firmware updates on the BCTRL with
a \texttt{canFwUpdate} flag (address 0x04CB) set to \texttt{false}. It changes the \texttt{HandleUART()} function (address 0x94C5) to accept or reject firmware update packets (starting with 0x07, 0x08, 0x09, or 0x0A), depending on \texttt{canFwUpdate}. This flag is set to \texttt{true} when the BCTRL receives our new unlock packet (0xEE followed by the unlock code).

\item[C2:] \emph{Manage UART Bus (MUB)}. MUB is used to send arbitrary UART packets on the internal bus.
It utilizes a USART1 peripheral mapped to RAM (address 0x5230). Due to V3, MUB crafts UART messages 
by altering fields such as the sender and receiver (e.g., 0x22 corresponds to BCTRL), packet type (e.g., read or write), command, payload, and CRC. It changes the \texttt{sendUART()} function (address 0xB06C) to send forged packets, stored from address 0xD3 onward. For example, \aphl\ uses MUB to spoof the BTS and retrieve the SHA256 hash of the e-scooter password from the DRV, but it can be used to spoof the DRV and BCTRL as well.

\item[C3:] \emph{Manage I2C Bus (MIB).} MIB is used to send custom packets to the BMON through the I2C bus. These packets are used to read and write registers on the BMON and control it. For example, it can be used to read battery cells voltages from \texttt{VC1}, \ldots, \texttt{VC10}
and alter the BMON undervoltage threshold in \texttt{UV\_TRIP}.
It rewrites \texttt{writeToI2C()} (address 0x90BE) to send forged packets on I2C.

\item[C4:] \emph{Modify Safety Thresholds (MST)}. MST alters the BMON undervoltage and overvoltage thresholds. It changes the BCTRL firmware boot logic, and initializes the BMON's \texttt{UV\_TRIP} (address 0x923D) and \texttt{OV\_TRIP} (address 0x923D) registers to \cuvt (lowest settable value) and \covt (highest settable value), or any other valid value. We remind that the default overvoltage and undervoltage safety thresholds are 2.75V and 4.2V. For example, MST is used in \aobda\ and \aubra\ to force an overvoltage or undervoltage state that physically damages the battery, while not triggering any BMON safety protection.

\item[C5:] \emph{Disable Safety Protections (DSP)}. DSP disables extra safety protections in the BMON, allowing critical overvoltage (above 4.7V) and undervoltage (below 1.58V). The patch sets to 0 the BMON's OV fault, UV fault, MOSFET charge, and MOSFET discharge bits in the \texttt{sys\_stat} register to prevent safety protections, such as preventing further charging, from triggering. It also rewrites the BCTRL error handler with NOPs to delete its OV and UV fault checks (address 0xB5A9) and ignore any fault message received from the BMON. For example, we use DSP in \aobda\ to disable overvoltage protections in order to damage the battery via critical overvoltage.

\item[C6:] \emph{Disable Load Balancing (DLB)}. Load balancing is essential (in any BES) to compensate for voltage discrepancies across cells. DLB disables it to speed up cell deterioration. It rewrites \seqsplit{\texttt{manageLoadBalancing()}} (address 0xA81A) to unset the three \texttt{CELLBAL} registers regulating balancing. For example, DLB is used by \aubra\ to forcefully imbalance and kill one specific cell at a time.

\item[C7:] \emph{Disable Battery Charging (DBC)}. DBC disallows charging by rewriting \texttt{controlCharge()} (address 0x945C) and setting \texttt{canCharge} (address 0x42C) to \texttt{false}, regardless if the charger is connected. For instance, DES5 prevents the battery from charging using DBC.

\item[C8:] \emph{Fast Battery Discharge (FBD)}. FBD accelerates battery consumption by altering the BCTRL main loop. It sets to \texttt{false} the \texttt{sleepMode} variable (address 0x400) and forces the BCTRL to stay in a more resource-intensive mode. FBD can also turn on the e-scooter's headlight and taillight to further increase battery consumption. \aubra\ uses FBD to accelerate battery depletion and critical undervoltage.

\item[C9:] \emph{Change BLE Advertising (CBA)}. CBA changes the e-scooter's BLE advertisements, including its BLE device name. It sends a proprietary message to the BTS containing the new BLE advertisement data. The message comprises a destination field (0x20), an opcode (0x50), and a payload containing the new advertisement. Once the message is received by the BTS, the e-scooter reboots and starts advertising over BLE with the new data. In \autia, the attacker tracks the e-scooter and its user via a custom BLE advertisement containing their fingerprint using CBA.

\item[C10:] \emph{Spoof Cell Voltage (SCV)}. SCV spoofs voltage measurements of battery cells to Mi Home to add stealthiness to an attack. It rewrites the BCTRL main loop to ignore the actual measurements from the BMON and reports arbitrary spoofed values. SCV sends these values to the BTS, which in turn sends them to Mi Home, which displays them to the user. By faking voltage, SCV allows \aubra\ to hide the real voltage from the user while keeping undervolting the battery cells under the hood.

\item[C11:] \emph{Alter Brake Settings (ABS)}. ABS alters brake-related settings managed by the BCTRL. It rewrites the default values for the brake lever pull (liner response between pull and braking force), brake phase current (regulates braking force), regenerative braking activation, and current raising coefficient (braking force in relation to speed). ABS can modify brake values in real-time depending on the context (e.g., downhill or crossing a road), maximizing the chances of injuring the user.

\end{itemize}

\begin{table}[tb]
  \caption{Mapping between the \aname\ attacks and the \caps\ capabilities offered by the \toolkit\ binary patcher (\texttt{payload-patcher.py)}. \cmark*: \aobda\ and \aubra\ use either C4 or C5, depending on the attack scenario.}
  \renewcommand{\arraystretch}{1.1}
  \centering\small
    \begin{tabular}{@{}lccccc@{}}
	    \toprule
	    \textbf{BCTRL Capability} & \textbf{\aobda} & \textbf{\aubra} & \textbf{\autia} & \textbf{\adesa} & \textbf{\aphla} \\
	    \midrule
	    C1: Disable Fw Updates & \cmark\ & \cmark\ & \cmark\ & \cmark\ & \cmark\ \\
	    C2: Manage UART Bus & \cmark\ & \cmark\ & \cmark\ & \cmark\ & \cmark\ \\
	    C3: Manage I2C Bus & \cmark\ & \cmark\ & \cmark\ & \cmark\ & \xmark\ \\
	    C4: Modify Safety Thr. & \hspace{1.4mm}\cmark* & \hspace{1.4mm}\cmark* & \xmark\ & \xmark\ & \xmark\ \\
	    C5: Disable Safety Prot. & \hspace{1.4mm}\cmark* & \hspace{1.4mm}\cmark* & \xmark\ & \xmark\ & \xmark\ \\
	    C6: Disable Load Bal. & \cmark\ & \cmark\ & \xmark\ & \xmark\ & \xmark\ \\
	    C7: Disable Batt. Charge & \xmark\ & \cmark\ & \xmark\ & \cmark\ & \xmark\ \\
	    C8: Fast Batt. Discharge & \xmark\ & \cmark\ & \xmark\ & \cmark\ & \xmark\ \\
	    C9: Change BLE Adv. & \xmark\ & \cmark\ & \cmark\ & \xmark\ & \cmark\ \\
	    C10: Spoof Cell Voltage & \cmark\ & \cmark\ & \xmark\ & \xmark\ & \xmark\ \\
	    C11: Alter Brake Settings & \xmark\ & \xmark\ & \xmark\ & \cmark\ & \xmark\ \\
	    \bottomrule
    \end{tabular}
    \label{tab:caps}
\end{table}

\textbf{Usage of C1--C11}.
Table~\ref{tab:caps} shows how we combine the eleven capabilities to perform \aubra, \aobda, \autia, \adesa, and \aphla. DFU is used in all attacks to prevent users from replacing our malicious BCTRL via BLE firmware updates. We use MUB and MIB in all attacks except MIB in \aphla, as they allow for controlling the UART and I2C buses. \aubra\ and \aobda\ use DLB, and either MST or DSP, to break safety mechanisms and induce undervoltage and overvoltage. MST prevents the BMON from activating safety protections and raising errors, resulting in an attack undetectable by the BMON, while DSP turns off safety protections as soon as they are activated, allowing for critical overvoltage and undervoltage. DBC is used in \aubra\ and DES5, as they are needed to target the e-scooter charger. FBD is used in \aubra\ and DES7, as it accelerates battery consumption. \aubra, \autia, and \aphla\ rely on CB to exfiltrate data from the e-scooter. \aubra\ and \aobda\ use SCV to tamper with battery readings. DES8 utilizes C11 to change downhill acceleration and mess with brake responsiveness.

\subsection{\aubra\ Ransom Android App and Backend}\label{subsec:impl-ra}

Our toolkit contains an Android app and a Django/MongoDB backend to test the \aubra\ ransomware in a controlled, but realistic environment. These components generate unique ransom unlock codes, manage cryptocurrency payments, and handle BCTRL recovery.

\textbf{ScooterToolkit}. We developed \emph{ScooterToolkit}, an Android app given to the victim so that they can pay the ransom in cryptocurrency and stop the undervoltage attack on their battery. The app is written in Kotlin and
requires Bluetooth, Location, and Internet permissions.
It initiates the BCTRL recovery when the back-end returns the unlock code
that enables firmware updates (i.e., the user paid the ransom).

\textbf{Django and MongoDB Backend}. Using Django, we developed a RESTful web service that exposes two APIs to the ransom app through the APIView class. The \texttt{simulatePayment()} API simulates a user payment, setting the payment status to \texttt{Paid}. The \texttt{\seqsplit{unlockFirmware(serial)}} API requires the e-scooter's motor serial as its input, and, if the ransom has been paid, the unlock code is returned. We also created a model serializer that converts e-scooter objects into JSON to interface with our MongoDB database. This database stores the e-scooter's motor serial and model, 128-bit ransomware unlock codes,
and payment status for each compromised device.

\section{Evaluation}\label{sec:eval}

We evaluated the \aname\ attacks (presented in Section~\ref{sec:attacks}) by using the \toolkit\ toolkit (described in Section~\ref{sec:impl}) on M365 and Mi3 e-scooters with up-to-date BTS, DRV, and BCTRL firmware. At the time of our experiments (2023), the M365 was the most popular model, while the Mi3 was Xiaomi's latest e-scooter. We also provide a video demonstration of UTI and DES4 at~\cite{etrojans-demo}. Next, we describe our attack setup and results.

\subsection{\aname\ Attacks Setup}\label{subsec:setup}

We connected to the victim's e-scooter and gained authorized access (e.g., via malicious pairing or session downgrade).
Next, we performed a rogue firmware update from proximity (within the e-scooter's BLE range) or remotely, through a malicious app installed on the victim's smartphone. We gained authorized access. In the proximity attack scenario, we utilized the \texttt{payload-patcher.py} script from a Dell Inspiron 15 3000 running a Linux-based OS. In the remote scenario, we used a Realme GT (Android 13) and a OnePlus 3 (Android 9), running the malicious \emph{ScooterToolkit} Android app. Both the script and the app can be found in our toolkit. We then set up each attack differently.

\textbf{\aobda\ Setup}. To induce overvoltage, we connected two 30V power supplies in series and set them up to output 50V (i.e., above the 4.7V maximum overvoltage threshold). We monitored the voltage, temperature, and other battery parameters directly from the BMON by using our dynamic debugging setup. We planned to stop the experiment when a cell reached either 50V or 80°C, to guarantee our safety (see Section~\ref{subsec:safety}).

\textbf{\aubra\ Setup}. We configured \aubra\ to activate when the e-scooter battery is charging at 5\% or lower. We then discharged the battery to that level and connected the e-scooter to a charger, simulating a user who recharges their device overnight. Using our dynamic debugging setup, we observed and timed the battery dropping from 5\% charge to 0\% (3.6V), and going beyond the minimum undervoltage threshold (below 1.58V). Since we wanted to assess how much permanent damage could be dealt, we maintained the battery in an undervolted state while monitoring parameters such as voltage, current, and temperature.

\textbf{\autia\ Setup}. We deployed BLE sniffers in four separate locations inside a private parking lot owned by our institution. These e-scooters traveled on a predetermined route along the four locations, while the sniffers logged their passing and the live data leaked by their BLE advertisements. The logs were then fed to a tracking system, updating the users' movements in real-time, and processed by scripts that attempted to de-anonymize them, based on where and for how long the users stopped during the current trip.

\textbf{\adesa\ Setup}. For each DES attack, we measured the time-to-failure and effect on user safety by observing their behaviour after the attack was triggered. We tested DES8 on a mild downward slope while wearing protective equipment to mitigate fall injuries.

\textbf{\aphla\ Setup}. We tested the feasibility of retrieving the six-digit e-scooter password when deploying \aphla. We considered the seven most common six-digit patterns~\cite{insecure-pins} and randomly generated seventy Xiaomi-compliant passwords (ten for each pattern). Then, we implemented a rainbow table~\cite{rainbow} and populated it with 3 million randomly generated Xiaomi-compliant passwords, simulating a basic attacker setup.

\subsection{\aname\ Attacks Results}\label{subsec:results}

\begin{table}[tb]
\caption{Evaluation results. All \attacks\ attacks are effective on the M365 and Mi3.\\
\cmark$^*$: \aubra\ with dangerous undervoltage (1.58V).}
\renewcommand{\arraystretch}{1.1}
\centering\small
\begin{tabular}{@{}lccl@{}}
	\toprule
	\textbf{Attack} & \textbf{M365} & \textbf{Mi3} & \textbf{Details} \\
	\midrule
       \aobda\ & \cmark\ & \cmark\ & Mi3 battery overheating in 5min \\
	\aubra\ & \cmark\ & \hspace{1mm}\cmark$^*$ & M365 battery autonomy -50\% in 3.5h \\
	\autia\ & \cmark\ & \cmark\ & Tracks and de-anonymized the user \\
	\adesa\ & \cmark\ & \cmark\ & Sudden downhill deceleration \\
	\aphla\ & \cmark\ & \cmark\ & Leaks user password \\
\bottomrule
\end{tabular}
\label{tab:eval-results}
\end{table}

Table~\ref{tab:eval-results} shows our evaluation results. The M365 and Mi3 are vulnerable to the \attacks\ attacks. Overall, the results demonstrate that overvoltage and undervoltage are relevant and practical threats, as the attacker can turn off safety protections on the BMS by programming the BMON from the rogue BCTRL firmware. Moreover, an e-scooter can be tracked over BLE, its password hash can be leaked over BLE and brute forced offline, and the e-scooter's internals can be DoSed. We note that
our empirical results represent a \emph{lower bound} as other e-scooters and BESs might be vulnerable to our attacks. For example, the Xiaomi Pro e-scooter runs BCTRL v1.2.6 (like the M365), and the 1S, Essential, and Pro2 run BCTRL v1.4.1 (like the Mi3). Next, we provide attack-specific insights.

\textbf{\aobda\ Eval}.
\aobda\ overvolted and overheated the M365 battery within five minutes.
We surpassed the maximum overvoltage threshold and monitored the rising internal temperature of the battery.
When it reached the point of venting and electrolyte decomposition (i.e., 80°C), we halted the experiment.
On the Mi3, we could only reach a dangerous overvoltage condition,
as the DRV would power off the e-scooter when detecting critical overvoltage,
preventing us from significantly overheating the battery.

\textbf{\aubra\ Eval}.
\aubra\ irreversibly reduced the battery autonomy of the M365
by approximately 50\% in three hours and thirty minutes.
We forced a critical undervoltage state and stopped the experiment after a few cells
reached 0V (i.e., the most critical undervoltage condition).
These cells now have limited charging capacity (i.e., 75\%)
and discharge four times faster than healthy cells.

Similarly, we reduced the battery autonomy of the Mi3
by around 10\% in ten hours, spending six of them
depleting the battery to 0\% (three hours for the M365).
We could not drop below the minimum undervoltage threshold due to the DRV independently
measuring an abnormal voltage in the motor (RAM register with offset 0x47)
and raising \emph{Error 24 - Supply Voltage out of range}.
As such, the Mi3 is better protected than the M365 against \aubra,
even though it still sustained significant damage.

\textbf{\autia\ Eval}.
\autia\ successfully tracks users, exfiltrates their private data over BLE, and de-anonymizes them.
We could identify the two e-scooters from their fingerprint (e.g., motor serial number),
allowing us to track users across the four areas where we deployed the BLE sniffers.
Our tracking system, for example, reported up-to-date mileage (e.g., 0x01B1 represents 433km)
and battery charge (e.g., 0x2D represents 45\%), while identifying the locations
where the user stopped along a predefined route.
We confirmed that, by leaking the battery charge, total mileage, last travel time, last travel mileage,
and average speed, we could infer where they stopped, for how long,
and whether they recharged the e-scooter (e.g., at their home).
We also verified that the user is still trackable after factory resetting the e-scooter.

\textbf{\adesa\ Eval}.
All \adesa\ attacks render the e-scooter unresponsive and not drivable within milliseconds.
DES1 and DES3 cause the DRV to raise \emph{Error 21 - No Communication with BMS},
halting the motor and powering off the e-scooter after ten seconds.
DES2 cuts power from all internals, permanently powering off the e-scooter.
DES4 prevents battery charging regardless of the charger hardware or e-scooter status (e.g., powered on/off).

DES5 induces an endless lock state or reboot cycle, rendering the e-scooter unable to move or power off.
DES6 raises \emph{Error 23 - Internal BMS not activated}, which induces a constant beeping noise, and \emph{Error 24 - Supply Voltage out of range}, which powers off the e-scooter.
DES7 ignores all sleep cycles even when the e-scooter is off, thus accelerating battery consumption.
DES8 arbitrarily disables and re-enables regenerative braking, causing sudden and unpredictable deceleration and acceleration,
putting a downhill driver at risk of falling off and injuring themselves. It also makes it harder to initiate braking
by altering the brake lever responsiveness and modifies braking force by increasing or decreasing the brake current.
By setting these parameters incorrectly on purpose, the attacker can generate too high a current and destroy the BMS board.

The \adesa\ attacks can target a variety of internal components and deny them.
To restore the e-scooter to its original state, the user needs to flash the original BMON firmware
via debug ports without the option of an over-the-air firmware update.
This operation is complex, as it requires disassembling the entire battery compartment.

\textbf{\aphla\ Eval}.
\aphla\ leaked the e-scooter password hash and enabled offline brute forcing.
We sniffed three rotating BLE advertisements broadcasted by a compromised e-scooter,
each containing part of the hash.
Since the e-scooter resets after changing its advertisement,
we rotate it in rapid succession,
tricking the user into thinking a single reboot happened instead of three.
We also confirmed that retrieving the e-scooter password from the hash is feasible.
Our non-optimized rainbow table cracked the seventy passwords, following the most common
six-digit patterns, in less than one second on average.

\subsection{\aname\ Safety Concerns} \label{subsec:safety}

According to our experiments, OBD and UBR can be used to set a battery on fire and cause it to explode due to \emph{induced hazards of overvoltage and undervoltage, and stealthiness}. We recall that the attacks do not require physical access to the e-scooter, OBD does not require a custom e-scooter charger, and UBR does not require a charger. These aspects further increase the safety implications of the attacks. Next, we summarize the experimental evidence we collected to prove our claims.

During our UBR test, we undervolted the battery \emph{below the lowest BMS undervoltage threshold} (i.e., below 1.58V down to 0V). This dangerous condition is also complex for users to detect, as the BCTRL spoofs valid battery levels back to Mi Home and the user. During the OBD tests, we overvolted the battery \emph{above the highest BMS overvoltage threshold} (i.e., above  4.7V up to 4.9V). Moreover, the battery reached a temperature of 80°C, whereas the recommended temperature range is between 0 °C and 45°C.

\section{Countermeasures}\label{sec:counter}

We discuss how to address the \aname\ attacks by design with \emph{\counters\ fixes (F1--F4)}. Each fix addresses one of the \vulns\ vulnerabilities from Section~\ref{subsec:vulns}, i.e., F1, F2, and F3 fix \aubra, \autia, and \aphla, while F4 addresses \adesa.

\begin{itemize}

\item[F1:] \emph{\cone}.
The BCTRL firmware should be encrypted at rest (e.g., when stored by Mi Home) and in transit (e.g., during a BLE firmware update). We recommend reusing the custom Xiaomi implementation of the TEA block cipher already used to encrypt the DRV firmware (since v0.1.7).

\item[F2:] \emph{\ctwo}.
The BCTRL firmware should be digitally signed by Xiaomi and verified by the BCTRL to prevent rogue firmware updates. We recommend reusing ECDSA, which is FIPS-compliant~\cite{nist-dss} and already employed to sign/verify the firmware of the DRV (since DRV v0.1.7) and BTS (since BTS v1.5.2).

\item[F3:] \emph{\cthree}.
The UART communication should be encrypted, authenticated, and integrity-protected. We suggest using standard embedded protocols such as \emph{Secure Channel Protocol 03 (SCP03)}~\cite{gp-scp03}. SCP03 provides confidentiality, integrity, authenticity, and replay protection using a lightweight scheme based on pre-shared symmetric keys, authenticated encryption, and AES. Since SCP03 is transport-agnostic, it can be adapted to UART with low engineering effort by adding message framing and flow control.

\item[F4:] \emph{\cfour}.
The UART bus should also be protected against DoS attacks. We recommend implementing an efficient rate-limiting mechanism for UART, such as the leaky bucket algorithm~\cite{logothetis1994leaky}, where excess packets are discarded if the incoming packet rate exceeds the outgoing rate. Combining F4 with F3 enhances DoS protection, as F3 discards messages failing the integrity checks.

\end{itemize}

The fixes are \emph{practical} and \emph{backward-compatible}.
They introduce negligible overhead, as they leverage lightweight cryptography (e.g., TEA and SCP03), while providing needed security guarantees, including confidentiality, authentication, and DoS protection. They are backward-compatible with the firmware and internals of the e-scooters.

\section{Related Work}\label{sec:related}

\textbf{E-Scooters' Security and Privacy}.
Research on e-scooters has studied their proprietary protocols
and rental ecosystems, but no work has focused on their \emph{internals}.
A recent paper analyzed the Xiaomi e-scooter ecosystem,
identifying issues in the proprietary BLE protocols used by the e-scooter and Mi Home~\cite{casagrande23espoofer}. Researchers identified vulnerabilities in the Bird e-scooter sharing platform~\cite{bird-webapp}
and discovered MitM attacks on the rental e-scooters of the Lime company~\cite{mitm-rude}.
Other studies focused on the privacy of user-related data in e-scooter rental Android apps~\cite{scooter_wisec_2022}
and the safety risks of riding e-scooters~\cite{safetyScooter1}.
In 2019, researchers from Zimperium revealed flaws in the M365 locking system that could halt a running e-scooter~\cite{zimperium}. The same year, Lanrat found that a M365 e-scooter did not enforce authentication over BLE~\cite{m365-authbypass}.

\textbf{E-Scooters Modding}.
There is a vibrant e-scooter modding community around the ScooterHacking~\cite{website-scooterhacking} website. The community released useful tools such as ones to interact with Xiaomi e-scooters~\cite{github-scooterhacking,app-shu} and perform firmware updates~\cite{botox-fwpatcher}.
Other modding tools, such as~\cite{fwtoolkit-scooterhacking}, focus on e-scooter performance by, for example, increasing the maximum speed value. Our toolkit focuses on the previously unexplored BCTRL firmware and can add new malicious capabilities to the e-scooter BMS.

\textbf{Attacks on Battery}.
The likelihood of battery overheating, thermal runaway, overcharging, and mechanical damage in electric vehicles has been investigated in~\cite{batterySafetyRisk1,batterySafetyRisk2}.
Researchers have proposed a battery drain attack on gas and electric vehicles~\cite{cho2018killed}, an electromagnetic interference attack against automotive batteries~\cite{szakaly2023assault},
and a physical access rogue firmware update to a Tesla BMS~\cite{kiley-tesla}.
Theoretical vulnerabilities in MacBook batteries, capable of overheating or turning off the laptop, have been discussed in~\cite{miller2011battery}.
The \aname\ attacks are \emph{orthogonal} as they target e-scooters' battery and BMS.

\textbf{Attacks on Embedded Firmware}.
Several papers focused on malicious embedded firmware, but none discussed a malicious BCTRL firmware causing battery overvoltage and undervoltage, or tracking e-scooter users.
Researchers attacked embedded firmware on programmable logic control (PLC)~\cite{garcia2017hey},
object trackers~\cite{roth2022airtag}, printers~\cite{cui2013firmware}, mice~\cite{maskiewicz2014mouse}, OS management systems~\cite{ibdah2020dark}, mobile bootloaders~\cite{redini2017bootstomp}, network equipment~\cite{costin2014large}, botnets~\cite{antonakakis2017understanding} and industrial wrenches~\cite{bosch24ransom}.

\textbf{Attacks on other BES}.
Security research on BESs has identified vulnerabilities across various domains.
Automotive researchers exploited the insecure CAN bus by compromising an Electronic Control Unit (ECU)~\cite{koscher2010experimental}. Similar works have exposed wireless charging stations and EV infrastructure to attacks such as MitM, DoS, and privacy leakage~\cite{wu2021time, liu2022privacy, ni2023uncovering, sarieddine2024covert, baker2019evphy}. Studies on drones have revealed issues in firmware updates, mobile apps, and wireless traffic~\cite{drone-fwup, drone-app-vuln, drone-bebop, drone-ardrone}. 
Several IoT devices, such as fitness trackers~\cite{casagrande2022breakmi, classen2018anatomy, goyal2016mind}, vacuum robots~\cite{giese2025ecovacs, woot2019vacuums, giese2018dustcloud}, crypto wallets~\cite{hardwear2023wallet}, and smart doorbells~\cite{hardwear2023doorbell}, have been found vulnerable to a broad range of attacks, including spoofing, eavesdropping, MitM, privacy, and physical attacks.

\section{Conclusion}\label{sec:conc}

BES, like e-scooters and drones, are interesting attack targets as their exploitation can result in security, privacy, and safety issues. This paper focuses on e-scooters and their internals as they have received limited attention so far, mainly due to their complexity and proprietary nature.
We focus on two pervasive Xiaomi e-scooter models, the M365 and Mi3. We RE their BMS, DRV, BTS subsystems, and their UART and I2C communication channels. We uncover \vulns\ critical design flaws (V1--V4) enabling remote code execution on the BCTRL firmware via a rogue firmware update over BLE. We use this attack primitive to design \attacks\ novel attacks we call \aname. Our attacks are executed with an attacker-controlled device in BLE proximity to the e-scooter or via a malicious app installed on the victim's phone.

\aobda\ puts the battery in an overvoltage state by disabling the BMS overvoltage protections and damages it until destruction without being noticed by the end user.
\aubra\ disables undervoltage protections on the BMS and forces the battery into an undervoltage state, progressively and silently damaging it. Then, it asks the user for a ransom.
\autia\ generates a unique e-scooter fingerprint from the e-scooter internals and broadcasts it over BLE to track the e-scooter and its user. It can optionally exfiltrate other sensitive data. \adesa\ is a collection of attacks capable of DoS-ing the e-scooter's internal systems and communication in several ways.

We develop \toolkit, a new toolkit that enables conducting the \aname\ attacks and creating new ones using a binary patcher script supporting \caps\ malicious capabilities (C1--C10). We carried out the attacks on actual devices, proving their real-world impact and practicality. We discuss four fixes (F1--F4) to address the attacks at the design level in an efficient and backward-compatible way. We responsibly disclosed our contributions to Xiaomi, who acknowledged them, assigned us a CVE and the highest bounty for our category, and, most importantly, fixed the issues on newer e-scooter models.


\section*{Availability}

We provide our toolkit and attack demos at
{\color{blue}\url{https://github.com/Skiti/Etrojans}}.
We released it as open-source, but, for safety reasons, we redacted the code responsible for generating malicious BMS firmware, which took months of RE to develop.
The toolkit implements the \aname\ attacks and shows our full attack chain in a video demonstration.
It also contains a patched BCTRL firmware that blocks BMS firmware updates to prevent our attacks and our RE Ghidra projects for the M365 and Mi3 BMS firmware.
Additionally, we release the undervoltage battery ransomware proof of concept, including a modified e-scooter firmware,
an Android ransom payment app, and a Django/MongoDB backend managing the payment.

\bibliographystyle{plain}
\bibliography{biblio}

@misc{BQ76930,
    author = {Texas Instrument},
    year = {2022},
    title = {{6 to 10-Series Cell Li-Ion and Li-Phosphate Battery Monitor}},
    howpublished ={\url{https://www.ti.com/product/BQ76930}},
}

@misc{BQ-datasheet,
    author = {Texas Instrument},
    year = {2022},
    title = {{BQ769x0 Datasheet}},
    howpublished ={\url{https://www.ti.com/lit/ds/symlink/bq76930.pdf}},
}

@misc{STM8,
    author = {STMicroelectronics},
    year = {2018},
    title = {{Ultra-low-power 8-bit MCU with 32 Kbytes Flash, 16 MHz CPU, integrated EEPROM}},
    howpublished ={\url{https://www.st.com/en/microcontrollers-microprocessors/stm8l151k6.html}},
}

@misc{stm8-migration-stm32,
    author = {STMicroelectronics},
    year = {2022},
    title = {{Migration of Applications from the STM8L and STM8S Series to the STM32C0 Series Microcontrollers}},
    howpublished ={\url{https://www.st.com/resource/en/application_note/an5775-migration-of-applications-from-the-stm8l-and-stm8s-series-to-the-stm32c0-series-microcontrollers-stmicroelectronics.pdf}},
}

@inproceedings{casagrande23espoofer,
    author = {Casagrande, Marco and Cestaro, Riccardo and Losiouk, Eleonora and Conti, Mauro and Antonioli, Daniele},
    title = {{E-Spoofer: Attacking and Defending Xiaomi Electric Scooter Ecosystem}},
    year = {2023},
    publisher = {Association for Computing Machinery},
    address = {New York, NY, USA},
    booktitle = {Proceedings of the 16th ACM Conference on Security and Privacy in Wireless and Mobile Networks},
    location = {Guildford, United Kingdom},
    series = {WiSec '23}
}

@misc{esp-gh,
    author = {Marco Casagrande},
    year = {2024},
    title = {{E-Spoofer (GitHub)}},
    howpublished ={\url{https://github.com/Skiti/ESpoofer}},
}

@misc{robo-fw,
    author = {Daljeet Nandha},
    year = {2022},
    title = {{Exploring Xiaomi’s New Firmware Security Measures}},
    howpublished ={\url{https://robocoffee.de/?p=193}},
}

@misc{gp-scp03,
    author = {Global Platform},
    year = {2019},
    title = {{Secure Channel Protocol 03}},
    howpublished ={\url{https://globalplatform.org/wp-content/uploads/2014/07/GPC\_2.3\_D\_SCP03\_v1.1.2\_PublicRelease.pdf}},
}

@misc{ti-ocuv,
    author = {Texas Instruments},
    year = {2020},
    title = {{How To Protect 48-V Batteries from Overcurrent and Undervoltage}},
    howpublished ={\url{https://www.ti.com/lit/an/snoaa65/snoaa65.pdf}},
}

@misc{xiaomi-trust-center,
    author = {Xiaomi},
    year = {2025},
    title = {{Xiaomi – Trust Center EOL Product List}},
    howpublished ={\url{https://trust.mi.com/misrc/updates/iot}},
}

@misc{battery-ov,
    author = {Berkeley Lab},
    year = {2016},
    title = {{Overcharge Protection Prevents Exploding Lithium Ion Batteries IB-3263}},
    howpublished ={\url{https://ipo.lbl.gov/lbnl3263}},
}

@article{battery-uv,
  author = {Sripad, Shashank and Kulandaivel, Sekar and Pande, Vikram and Sekar, Vyas and Viswanathan, Venkatasubramanian},
  title = {{Vulnerabilities of Electric Vehicle Battery Packs to Cyberattacks on Auxiliary Components}},
  journal={arXiv preprint arXiv:1711.04822},
  year = {2017},
}

@article{mechanism-uv,
  author = {Guo, Rui and Ouyang, Minggao and Lu, Languang and Feng, Xuning},
  year = {2016},
  pages = {30248},
  title = {{Mechanism of the Entire Overdischarge Process and Overdischarge-induced Internal Short Circuit in Lithium-ion Batteries}},
  journal = {Scientific Reports},
}

@inproceedings{insecure-pins,
  author = {Wang, Ding and Gu, Qianchen and Huang, Xinyi and Wang, Ping},
  title = {Understanding Human-Chosen PINs: Characteristics, Distribution and Security},
  year = {2017},
  publisher = {Association for Computing Machinery},
  booktitle = {Proceedings of the 2017 ACM on Asia Conference on Computer and Communications Security (ASIA CCS'17)},
}

@misc{bev-market-size,
  title={Battery Electric Vehicles - Worldwide},
  author={Statista},
  year={2025},
  howpublished={\url{https://www.statista.com/outlook/mmo/electric-vehicles/electric-passenger-cars/battery-electric-vehicles/worldwide}},
}

@misc{escooter-market-size,
  title={{Electric Scooters Market Size, Share \& Trends Analysis Report (2024-2030)}},
  author={Grand View Research},
  year={2024},
  howpublished={\url{https://www.grandviewresearch.com/industry-analysis/electric-scooters-market}},
}

@misc{escooter-market-size2,
  title={Micro-mobility Market Size, Share \& Trends Analysis Report By Vehicle Type (Electric Kick Scooters, Electric Skateboards, Electric Bicycles), By Battery, By Voltage, By Region, And Segment Forecasts, 2025-2030},
  author={Grand View Research},
  year={2025},
  howpublished={\url{https://www.grandviewresearch.com/industry-analysis/micro-mobility-market-report}},
}

@misc{mihome-android,
  title={{Mi Home (Android)}},
  author={Xiaomi},
  year={2024},
  howpublished={\url{https://play.google.com/store/apps/details?id=com.xiaomi.smarthome}},
}

@misc{mihome-ios,
  title={{Mi Home (iOS)}},
  author={Xiaomi},
  year={2024},
  howpublished={\url{https://apps.apple.com/us/app/mi-home-xiaomi-smart-home/id957323480}},
}

@misc{xiaomi-bugbounty,
    title={{Xiaomi Bug Bounty Program (HackerOne)}},
    author={Xiaomi},
    year={2024},
    howpublished={\url{https://hackerone.com/xiaomi}},
}

@misc{xiaomi-leader,
    title={{Folding Electric Scooter Market: Current Analysis and Forecast (2023-2030)}},
    author={Univ Datos},
    year={2022},
    howpublished={\url{https://univdatos.com/report/folding-electric-scooter-market/}},
}

@misc{ninebot-sold,
    title={{Segway-Ninebot Eyes Bigger Market Share in Country}},
    author={Ma Si},
    year={2022},
    howpublished={\url{https://global.chinadaily.com.cn/a/202204/28/WS6269ee14a310fd2b29e59d1f.html}},
}

@inproceedings{wu2021time,
    title={{Time to rethink the design of Qi standard? security and privacy vulnerability analysis of Qi wireless charging}},
    author={Wu, Yi and Li, Zhuohang and Van Nostrand, Nicholas and Liu, Jian},
    booktitle={Annual Computer Security Applications Conference},
    pages={916--929},
    year={2021}
}

@inproceedings{ni2023uncovering,
    title={{Uncovering User Interactions on Smartphones via Contactless Wireless Charging Side Channels}},
    author={Ni, Tao and Zhang, Xiaokuan and Zuo, Chaoshun and Li, Jianfeng and Yan, Zhenyu and Wang, Wubing and Xu, Weitao and Luo, Xiapu and Zhao, Qingchuan},
    booktitle={2023 IEEE Symposium on Security and Privacy (SP)},
    pages={3399--3415},
    year={2023},
    organization={IEEE}
}

@article{liu2022privacy,
    title={Privacy leakage in wireless charging},
    author={Liu, Jianwei and Zou, Xiang and Zhao, Leqi and Tao, Yusheng and Hu, Sideng and Han, Jinsong and Ren, Kui},
    journal={IEEE Transactions on Dependable and Secure Computing},
    year={2022},
    publisher={IEEE}
}

@article{miller2011battery,
    title={Battery Firmware Hacking},
    author={Miller, Charlie},
    journal={Black Hat USA},
    pages={3--4},
    year={2011}
}

@article{cho2018killed,
    title={Who killed my parked car?},
    author={Cho, Kyong-Tak and Kim, Yuseung and Shin, Kang G},
    journal={arXiv preprint arXiv:1801.07741},
    year={2018}
}

@article{szakaly2023assault,
    title={{Assault and Battery: Evaluating the Security of Power Conversion Systems Against Electromagnetic Injection Attacks}},
    author={Szak{\'a}ly, Marcell and K{\"o}hler, Sebastian and Strohmeier, Martin and Martinovic, Ivan},
    journal={arXiv preprint arXiv:2305.06901},
    year={2023}
}

@inproceedings{redini2017bootstomp,
    title={BootStomp: on the security of bootloaders in mobile devices},
    author={Redini, Nilo and Machiry, Aravind and Das, Dipanjan and Fratantonio, Yanick and Bianchi, Antonio and Gustafson, Eric and Shoshitaishvili, Yan and Kruegel, Christopher and Vigna, Giovanni},
    booktitle={26th USENIX Security Symposium (USENIX Security 17)},
    pages={781--798},
    year={2017}
}

@inproceedings{ibdah2020dark,
    title={Dark firmware: a systematic approach to exploring application security risks in the presence of untrusted firmware},
    author={Ibdah, Duha and Lachtar, Nada and Elkhail, Abdulrahman Abu and Bacha, Anys and Malik, Hafiz},
    booktitle={23rd International Symposium on Research in Attacks, Intrusions and Defenses (RAID 2020)},
    pages={413--426},
    year={2020}
}

@inproceedings{maskiewicz2014mouse,
    title={{Mouse trap: Exploiting firmware updates in USB peripherals}},
    author={Maskiewicz, Jacob and Ellis, Benjamin and Mouradian, James and Shacham, Hovav},
    booktitle={8th USENIX Workshop on Offensive Technologies (WOOT 14)},
    year={2014}
}

@inproceedings{roth2022airtag,
    title={AirTag of the clones: shenanigans with liberated item finders},
    author={Roth, Thomas and Freyer, Fabian and Hollick, Matthias and Classen, Jiska},
    booktitle={2022 IEEE Security and Privacy Workshops (SPW)},
    pages={301--311},
    year={2022},
    organization={IEEE}
}

@inproceedings{garcia2017hey,
    title={{Hey, My Malware Knows Physics! Attacking PLCs with Physical Model Aware Rootkit}},
    author={Garcia, Luis and Brasser, Ferdinand and Cintuglu, Mehmet Hazar and Sadeghi, Ahmad-Reza and Mohammed, Osama A and Zonouz, Saman A},
    booktitle={NDSS},
    pages={1--15},
    year={2017}
}

@inproceedings{cui2013firmware,
    title={When firmware modifications attack: A case study of embedded exploitation},
    author={Cui, Ang and Costello, Michael and Stolfo, Salvatore},
    booktitle={NDSS},
    year={2013}
}

@misc{drone-app-vuln,
    title={{Multidimensional Attack Vectors and Countermeasures}},
    author={Luo, Aaron},
    year={2016},
    booktitle={DEF CON 24},
    howpublished={\url{https://media.defcon.org/DEF CON 24/DEF CON 24 presentations/DEF CON 24 - Aaron-Luo-Drones-Hijacking-Multi-Dimensional-Attack-Vectors-And-Countermeasures-UPDATED.pdf}},
}

@misc{drone-fwup,
    title={{Unchained Skies: A Deep Dive into Reverse Engineering and Exploitation of Drones}},
    author={Schloegel, Moritz and Schiller, Nico},
    year={2023},
    booktitle={REcon 2023},
    howpublished={\url{https://cfp.recon.cx/2023/talk/HLHH89/}},
}

@misc{drone-ardrone,
    title={{ARDrone Corruption}},
    author={Deligne, Eddy},
    year={2011},
    booktitle={Journal in Computer Virology},
    howpublished={\url{https://link.springer.com/article/10.1007/s11416-011-0158-4}},
}

@misc{drone-bebop,
    title={{Parrot Drones Hijacking}},
    author={Cabrera, Pedro},
    year={2018},
    booktitle={RSA Conference},
    howpublished={\url{https://www.rsaconference.com/Library/presentation/USA/2018/parrot-drones-hijacking}},
}

@misc{nist-dss,
    title={{FIPS 186-5 -- Digital Signature Standard (DSS)}},
    author={NIST},
    year={2023},
    howpublished={\url{https://nvlpubs.nist.gov/nistpubs/FIPS/NIST.FIPS.186-5.pdf}},
}

@misc{zimperium,
  title={{Don't Give Me A Brake – Xiaomi Scooter Hack Enables Dangerous Accelerations And Stops For Unsuspecting Riders}},
 	author={Rani Idan (Zimperium)},
 	year={2019},
  howpublished={\url{https://blog.zimperium.com/dont-give-me-a-brake-xiaomi-scooter-hack-enables-dangerous-accelerations-and-stops-for-unsuspecting-riders/}},
}

@misc{m365-authbypass,
  title={{Xiaomi M365 Scooter Authentication Bypass}},
	author={Ian D. Foster},
 	year={2019},
  howpublished={\url{https://lanrat.com/xiaomi-m365/}},
}

@misc{bird-webapp,
  title={{App Analysis: Bird}},
  author={App Analyst},
  year={2019},
  howpublished={\url{https://theappanalyst.com/bird.html/}},
}

@misc{bird-equals-m365,
  title={{Security Engineering: Inside the Scooter Startups}},
  author={Brian Benchoff},
  year={2019},
  howpublished={\url{https://hackaday.com/2019/02/12/security-engineering-inside-the-scooter-startups/}},
}

@misc{who-makes-bird,
  title={{Who Makes Bird and Lime Scooters?}},
  author={Tech We Want},
  year={2018},
  howpublished={\url{https://techwewant.com/this-is-who-makes-bird-lime-and-jump-scooters-review-b3f6be32221e}},
}

@misc{mitm-rude,
  title={{Scooters Hacked To Play Rude Messages To Riders}},
  author={BBC News},
  year={2019},
  howpublished={\url{https://www.bbc.com/news/technology-48065432}},
}

@inproceedings{scooter_wisec_2022,
author = {Vinayaga-Sureshkanth, Nisha and Wijewickrama, Raveen and Maiti, Anindya and Jadliwala, Murtuza},
title = {An Investigative Study On The Privacy Implications Of Mobile E-Scooter Rental Apps},
year = {2022},
booktitle = {Proceedings of the 15th ACM Conference on Security and Privacy in Wireless and Mobile Networks},
pages = {125–139},
}

@inproceedings{isik2023platforms,
  author={Isik, Gizay Kisa and Tryfonas, Theo and Oikonomou, George},
  booktitle={2023 IEEE 26th International Conference on Intelligent Transportation Systems (ITSC)}, 
  title={E-scooter Sharing Platforms: Understanding Their Architecture and Cybersecurity Threats}, 
  year={2023},
  pages={5909-5916}
}

@misc{cameron2019iot,
  title={IoT Penetration Testing: Hacking an Electric Scooter},
  author={Cameron Booth, Louis and Mayrany, Matay},
  year={2019}
}

@article{safetyScooter1,
title = {E-Scooter safety: The riding risk analysis based on mobile sensing data},
journal = {Accident Analysis and Prevention},
year = {2021},
author = {Qingyu Ma and Hong Yang and Alan Mayhue and Yunlong Sun and Zhitong Huang and Yifang Ma},
}

@article{batterySafetyRisk1,
title = {Full-scale experimental study on suppressing lithium-ion battery pack fires from electric vehicles},
journal = {Fire Safety Journal},
pages = {103562},
year = {2022},
author = {Yan Cui and Jianghong Liu and Xin Han and Shaohua Sun and Beihua Cong},
}

@article{batterySafetyRisk2,
title = {Safety issues and mechanisms of lithium-ion battery cell upon mechanical abusive loading: A review},
journal = {Energy Storage Materials},
pages = {85-112},
year = {2020},
author = {Binghe Liu and Yikai Jia and Chunhao Yuan and Lubing Wang and Xiang Gao and Sha Yin and Jun Xu},
}

@misc{website-scooterhacking,
  title={{ScooterHacking Website}},
  author={ScooterHacking},
  year={2022},
  howpublished={\url{https://scooterhacking.org/}},
}

@misc{github-scooterhacking,
  title={{ScooterHacking (GitHub)}},
  author={ScooterHacking},
  year={2022},
  howpublished={\url{https://github.com/orgs/scooterhacking/repositories}},
}

@misc{app-shu,
  title={{ScooterHacking Utility App}},
  author={ScooterHacking},
  year={2022},
  howpublished={\url{https://play.google.com/store/apps/details?id=sh.cfw.utility}},
}

@misc{home-shu,
  title={{ScooterHacking Utility Homepage}},
  author={ScooterHacking},
  year={2023},
  howpublished={\url{https://utility.cfw.sh/}},
}

@misc{fwtoolkit-scooterhacking,
  title={{ScooterHacking Firmware Toolkit}},
  author={ScooterHacking},
  year={2022},
  howpublished={\url{https://mi.cfw.sh/}},
}

@misc{botox-fwpatcher,
  title={{Xiaomi M365 Firmware Patcher}},
  author={BotoX},
  year={2022},
  howpublished={\url{https://github.com/BotoX/xiaomi-m365-firmware-patcher}},
}

@misc{rainbow,
    author = {Lu, Charles and Ding, Jialin},
    year = {2015},
    title = {{Rainbow Table Attack on 6-Digit PINs (GitHub)}},
    howpublished ={\url{https://github.com/clu8/RainbowTable}},
}

@misc{botox-tea,
    author = {BotoX},
    year = {2024},
    title = {{Xiaomi TEA Implementation (GitHub)}},
    howpublished ={\url{https://github.com/BotoX/xiaomi-m365-firmware-patcher/blob/master/xiaotea/xiaotea.py}},
}

@misc{kiley-tesla,
  title={{Reverse Engineering the Tesla Battery Management System to Increase Power Available}},
  author={Patrick Kiley},
  year={2020},
  howpublished={\url{https://www.youtube.com/watch?v=UV2zvgyIF0I}},
}

@inproceedings{uher2016denial,
    title={{Denial of Sleep Attacks in Bluetooth Low Energy Wireless Sensor Networks}},
    author={Uher, Jason and Mennecke, Ryan G and Farroha, Bassam S},
    booktitle={MILCOM 2016-2016 IEEE Military Communications Conference},
    pages={1231--1236},
    year={2016},
    organization={IEEE}
}

@misc{bosch24ransom,
  title={{Network-Connected Torque Wrench Used in Factories Is Vulnerable to Ransomware}},
  author={PCmag},
  year={2024},
  howpublished={\url{https://www.pcmag.com/news/network-connected-torque-wrench-used-in-factories-is-vulnerable-to-ransomware}},
}

@misc{pagers,
  title={{Israel concealed explosives inside batteries of pagers sold to Hezbollah, Lebanese officials say}},
  author={CNN},
  year={2024},
  howpublished = {\url{https://edition.cnn.com/2024/09/27/middleeast/israel-pager-attack-hezbollah-lebanon-invs-intl/index.html}},
}

@inproceedings{costin2014large,
    title={{A large-scale analysis of the security of embedded firmwares}},
    author={Costin, Andrei and Zaddach, Jonas and Francillon, Aur{\'e}lien and Balzarotti, Davide},
    booktitle={23rd USENIX security symposium (USENIX Security 14)},
    pages={95--110},
    year={2014}
}

@inproceedings{antonakakis2017understanding,
    title={Understanding the mirai botnet},
    author={Antonakakis, Manos and April, Tim and Bailey, Michael and Bernhard, Matt and Bursztein, Elie and Cochran, Jaime and Durumeric, Zakir and Halderman, J Alex and Invernizzi, Luca and Kallitsis, Michalis and others},
    booktitle={26th USENIX security symposium (USENIX Security 17)},
    pages={1093--1110},
    year={2017}
}

@inproceedings{koscher2010experimental,
    title={Experimental security analysis of a modern automobile},
    author={Koscher, Karl and Czeskis, Alexei and Roesner, Franziska and Patel, Shwetak and Kohno, Tadayoshi and Checkoway, Stephen and McCoy, Damon and Kantor, Brian and Anderson, Danny and Shacham, Hovav and others},
    booktitle={2010 IEEE symposium on security and privacy},
    pages={447--462},
    year={2010},
    organization={IEEE}
}

@inproceedings{checkoway2011comprehensive,
    title={Comprehensive experimental analyses of automotive attack surfaces},
    author={Checkoway, Stephen and McCoy, Damon and Kantor, Brian and Anderson, Danny and Shacham, Hovav and Savage, Stefan and Koscher, Karl and Czeskis, Alexei and Roesner, Franziska and Kohno, Tadayoshi},
    booktitle={20th USENIX security symposium (USENIX Security 11)},
    year={2011}
}

@inproceedings{wolf2004security,
  title={Security in automotive bus systems},
  author={Wolf, Marko and Weimerskirch, Andr{\'e} and Paar, Christof},
  booktitle={Workshop on Embedded Security in Cars},
  pages={1--13},
  year={2004},
  organization={Bochum}
}

@inproceedings{markert2020pin,
    title={This pin can be easily guessed: Analyzing the security of smartphone unlock pins},
    author={Markert, Philipp and Bailey, Daniel V and Golla, Maximilian and D{\"u}rmuth, Markus and Aviv, Adam J},
    booktitle={2020 IEEE Symposium on Security and Privacy (SP)},
    pages={286--303},
    year={2020},
    organization={IEEE}
}

@inproceedings{sarieddine2024covert,
  author = {Sarieddine, Khaled and Sayed, Mohammad Ali and Torabi, Sadegh and Attallah, Ribal and Jafarigiv, Danial and Assi, Chadi and Debbabi, Mourad},
  title = {Uncovering Covert Attacks on EV Charging Infrastructure: How OCPP Backend Vulnerabilities Could Compromise Your System},
  year = {2024},
  booktitle = {Proceedings of the 19th ACM Asia Conference on Computer and Communications Security},
  pages = {977–989},
  numpages = {13},
}

@inproceedings{baker2019evphy,
  author = {Baker, Richard and Martinovic, Ivan},
  title = {Losing the Car Keys: Wireless {PHY-Layer} Insecurity in {EV} Charging},
  booktitle = {28th USENIX Security Symposium (USENIX Security 19)},
  year = {2019},
  pages = {407--424},
}

@article{classen2018anatomy,
    title={{Anatomy of a Vulnerable Fitness Tracking System: Dissecting the Fitbit Cloud, App, and Firmware}},
    author={Jiska Classen and Daniel Wegemer and Paul Patras and Tom Spink and Matthias Hollick},
    journal={Proceedings of the ACM on Interactive, Mobile, Wearable and Ubiquitous Technologies (IMWUT '18)},
    volume={2},
    number={1},
    pages={1--24},
    year={2018},
}

@article{casagrande2022breakmi, 
	title={BreakMi: Reversing, Exploiting And Fixing Xiaomi Fitness Tracking Ecosystem}, 
	volume={2022}, 
	number={3}, 
	journal={IACR Transactions On Cryptographic Hardware And Embedded Systems}, 
	author={Casagrande, Marco and Losiouk, Eleonora and Conti, Mauro and Payer, Mathias and Antonioli, Daniele}, year={2022}, 
	pages={330–366}
}

@inproceedings{goyal2016mind,
	author = {Rohit Goyal and Nicola Dragoni and Angelo Spognardi},
	title = {{Mind the Tracker You Wear: A Security Analysis of Wearable Health Trackers}},
	year = {2016},
	booktitle = {Proceedings of the 31st Annual ACM Symposium on Applied Computing (SAC '16), Pisa, Italy},
	pages = {131–136},
	numpages = {6},
}

@misc{giese2025ecovacs,
    author = {Dennis Giese and Braelynn},
    year = {2025},
    title = {{DEFCON 32 - Reverse engineering and hacking Ecovacs robots}},
    howpublished ={\url{https://youtu.be/_wUsM0Mlenc?si=ACJj36Xvic03okKj}},
}

@misc{giese2018dustcloud,
    title={{Having fun with IoT: Reverse Engineering and Hacking of Xiaomi IoT Devices}},
    author={Dennis Giese},
    year={2018},
    howpublished ={\url{https://dontvacuum.me/talks/DEFCON26/DEFCON26-Having_fun_with_IoT-Xiaomi.html}},
}

@inproceedings{woot2019vacuums,
	author = {Fabian Ullrich and Jiska Classen and Johannes Eger and Matthias Hollick},
	title = {Vacuums In Ihe Cloud: Analyzing Security In A Hardened IoT Ecosystem},
	booktitle = {13th USENIX Workshop on Offensive Technologies (WOOT 19)},
	year = {2019},
	url = {https://www.usenix.org/conference/woot19/presentation/ullrich},
}

@misc{hardwear2023wallet,
    author = {Michael Mouchous and Karim Abdellatif},
    year = {2023},
    title = {{Hardwear.io USA - The Misuse Of Secure Components In Hardware Wallets}},
    howpublished ={\url{https://www.youtube.com/watch?v=BoOTVeFL30I}},
}

@misc{hardwear2023doorbell,
    author = {Daniel Schwendner},
    year = {2023},
    title = {{Hardwear.io NL - Hacking A Smart Doorbell: An IoT Hacking Guide}},
    howpublished ={\url{http://youtube.com/watch?v=pfp5IaQyIkg}},
}

@misc{etrojans-demo,
    author = {E-Trojans},
    year = {2024},
    title = {{User Tracking and DoS demonstration on the Xiaomi Mi3 e-scooter}},
    howpublished ={\url{https://www.youtube.com/watch?v=Y2R46yeCXOQ}},
}

@article{polarity-inversion,
	author = {Menale, Carla and Constà, Stefano and Sglavo, Vincenzo and Della Seta, Livia and Bubbico, Roberto},
	title = {{Experimental Investigation of Overdischarge Effects on Commercial Li-Ion Cells}},
	journal = {Energies},
	volume = {15},
	year = {2022},
	number = {22},
}

@article{overheat-stability,
	author = {Oltean, Gabriel and Plylahan, Nareerat and Ihrfors, Charlotte and Wei, Wei and Xu, Chao and Edström, Kristina and Nyholm, Leif and Johansson, Patrik and Gustafsson, Torbjörn},
	title = {{Towards Li-Ion Batteries Operating at 80 °C: Ionic Liquid versus Conventional Liquid Electrolytes}},
	journal = {Batteries},
	volume = {4},
	year = {2018},
	number = {1},
}

@inproceedings{logothetis1994leaky,
  title={{The Leaky Bucket as a Policing Device: Transient Analysis and Dimensioning}},
  author={Logothetis, Dimitris and Trivedi, K},
  booktitle={INFOCOM},
  volume={94},
  year={1994}
}


\end{document}